\documentclass[a4paper,fleqn]{cas-dc}

\usepackage[authoryear,longnamesfirst]{natbib}

\usepackage{upgreek}

\usepackage{amsmath,amsfonts,amssymb}
\usepackage{algorithm}
\usepackage{algpseudocode}

\usepackage{pgfplots}
\pgfplotsset{compat=1.18} 

\usepackage{adjustbox}

\usepackage{xcolor}

\def\tsc#1{\csdef{#1}{\textsc{\lowercase{#1}}\xspace}}
\tsc{WGM}
\tsc{QE}
\begin{document}
\let\WriteBookmarks\relax
\def\floatpagepagefraction{1}
\def\textpagefraction{.001}

\shorttitle{Primal-Dual Formulation for Pricing Voltage Stability Services}    

\shortauthors{Peng Wang, Luis Badesa}  

\title [mode = title]{A Primal-Dual Formulation for Pricing Static Voltage Stability Services within a Unit Commitment Model}  




\author[1]{Peng Wang}%[<options>]

%\cormark[1]


\ead{peng.wang@alumnos.upm.es}



\affiliation[1]{organization={School of Industrial Engineering and Design (ETSIDI), Technical University of Madrid (UPM)},
            city={Madrid},
%          citysep={}, % Uncomment if no comma needed between city and postcode
            %postcode={}, 
            state={Community of Madrid},
            country={Spain}}

\author[1]{Luis Badesa}%[]
\cormark[1]

% Footnote of the second author

% Email id of the second author
\ead{luis.badesa@upm.es}



%\affiliation[2]{organization={},
%          citysep={}, % Uncomment if no comma needed between city and postcode

\cortext[1]{Corresponding author}

\fntext[1]{https://orcid.org/0009-0008-0658-8649}
\fntext[2]{https://orcid.org/0000-0003-2411-3061}

 \nonumnote{This work was supported by MICIU/AEI/10.13039/501100011033 and ERDF/EU under grant PID2023-150401OA-C22, as well as by the Madrid Government (Comunidad de Madrid-Spain) under the Multiannual Agreement 2023-2026 with Universidad Politécnica de Madrid, ``Line A - Emerging PIs'' (grant number: 24-DWGG5L-33-SMHGZ1). The work of Peng Wang was also supported by China Scholarship Council under grant 202408500065.}

\begin{abstract}
In modern power systems with high penetration of Inverter-Based Resources (IBR), most converters operate in Grid-Following (GFL) mode. Some buses exhibit inherently low Short-Circuit Ratios (SCRs), a property majorly shaped by network topology. The integration of GFL-IBR onto such weak buses thus demands attention to static voltage stability. To address this issue, market mechanisms have been proposed to incentivize generators to provide voltage stability services, such as commitment of synchronous generators for reducing the equivalent impedance at low SCR buses and adaptive reactive power support from GFL-IBR. To compute shadow prices for these services, previously proposed methods such as the `restricted' and `dispatchable' approaches may fail to guarantee operating cost recovery for voltage-stability service providers. As the resulting prices are determined purely from a social surplus maximization objective, the profitability of units is entirely overlooked. This suggests that new pricing methodologies are needed to satisfy cost-recovery requirements. Therefore, this paper proposes a pricing method based on a primal-dual formulation. Case studies demonstrate that the proposed method can consistently produce revenue-adequate shadow prices, enabling all participating units to recover their costs without supplementary uplift payments.
\end{abstract}


\begin{highlights}
\item A primal‑dual pricing formulation is proposed to compute shadow prices for static voltage stability services under mixed‑integer second‑order‑cone unit‑commitment constraints, preserving binary commitment decisions of synchronous generators.
\item Explicit non‑negative‑profit constraints incorporating both short‑run operating costs and daily‑amortized investment costs of inverter‑based resources guarantee revenue adequacy for synchronous generators, VSGs and GFL‑IBR simultaneously.
\item The derived shadow prices deliver sufficient investment incentives for voltage‑supporting resources and eliminate the need for unit‑specific uplift/make‑whole payments, overcoming transparency and cost‑allocation drawbacks of restricted and dispatchable pricing schemes.
\end{highlights}

%\nocite{*}

% Keywords
\begin{keywords}
 Inverter-based resources, static voltage stability services, shadow pricing, non-convexity, primal-dual formulation
\end{keywords}

\maketitle

\section*{Nomenclature}
\addcontentsline{toc}{section}{Nomenclature}

\subsection*{Indices and Sets}
\begin{description}
    \item[$g_c,\mathcal{G}_c$] Index, Set of conventional SGs
    \item[$g_f, \mathcal{G}_f$] Index, Set of GFL-IBR  
    \item[$g_v,\mathcal{G}_v$]  Index, Set of VSGs
    \item[$m,\mathcal{M}$]  Index, Set for products of units' operating states
    \item[$t,T$]  Index, Set of operation time periods
    \item[$\Phi(\cdot)$]  Index for buses of generators
\end{description}

\subsection*{Constants and Parameters}
\begin{description}
    \item[$\mathrm{c}_{g_c}^\mathrm{m}$]  Marginal generation costs of SGs (\texteuro/MWh)
    \item[$\mathrm{c}_{g_c}^\mathrm{nl}$]  No-load costs of SGs (\texteuro/h)
    \item[$\mathrm{c}^\mathrm{st}_{g_c}$]  Startup costs of SGs (\texteuro/h) 
    \item[$\mathrm{P}_i^{\mathrm{D}}, \mathrm{Q}_i^{\mathrm{D}}$]  \parbox[t]{.9\linewidth} {Active/reactive demand at bus $i$ (MW, Mvar)}
    \item[$\mathrm{P}_{\{\cdot\}}^{\mathrm{min}}, \mathrm{P}_{\{\cdot\}}^{\mathrm{max}}$]   \parbox[t]{.9\linewidth} {Generators' active power output limits (MW)}
    \item[$ \mathrm{Q}_{\{\cdot\}}^{\mathrm{min}}, \mathrm{Q}_{\{\cdot\}}^{\mathrm{max}}$]   \parbox[t]{.9\linewidth} {Generators' reactive power output limits (Mvar)}
    \item[$ \mathrm{S}^{\mathrm{max}}_{ \{g_c,g_f,g_v\} }$]  Rated apparent power of generators (MVA)
    \item[$\upalpha_{g_f},\upalpha_{g_v}$]  \parbox[t]{.9\linewidth} {Capacity percentage of GFL-IBR and VSGs}    
\end{description}

\subsection*{Variables}
\begin{description}
    \item[$C_{g_c}^\mathrm{st}$]  \parbox[t]{.9\linewidth} {Costs incurred by startup of SGs (\texteuro)}
    \item[$P_{\{\cdot\}}, Q_{\{\cdot\}}$]  \parbox[t]{.9\linewidth} {Generators' active/reactive output (MW, Mvar)}
    \item[$u_{g_c}$]  Binary variable, commitment of SGs
    \item[$\eta_m$]  Product of operating states of each two (V)SGs
    \item[$\lambda_{2,g_f},\mu_{g_f}$]  \parbox[t]{.8\linewidth} {Dual variables, associated with voltage stability constraints}
    \item[$\lambda_{g_c,\mathrm{commit}}$]  Dual variable, commitment prices (\texteuro/h)
     \item[$\lambda^{\mathrm{E}}$]  Dual variable, energy price (\texteuro/MWh)
\end{description}

\section{Introduction}
With the increasing penetration of Grid-following (GFL) Inverter-Based Resources (IBR) in low-carbon power systems, voltage stability at GFL-IBR buses has emerged as an essential operational concern \cite{zhou2024comparison}. In contrast to voltage-source Grid-forming (GFM) units, GFL-IBR act as current sources without inherent terminal voltage regulation. Furthermore, due to the remote, disperse placement of wind and solar renewable energy sources, these IBR are often connected to weak-grid regions with high impedance, which lowers the local Short-Circuit Ratio (SCR) and elevates voltage instability risks \cite{abdelwahab2026beyond}, thereby highlighting the need to explicitly ensure voltage stability at GFL-IBR buses.

Numerous methods have been developed to improve voltage stability. The deployment of Synchronous Condensers (SCs) is analyzed in \cite{bao2024maximizing} to enhance system stability by reducing the equivalent system impedance. Reference \cite{li2022grid} combines IBR with a co-located SC, allowing the unit to regulate its terminal voltage. To investigate the effect of SCs on system stability enhancement, \cite{hadavi2024quantifying} refines the SCR metric to enable a more comprehensive evaluation. However, installing new SCs to achieve the aforementioned system stability incurs additional investment costs; thus, \cite{fouladi2024optimal} determines an optimal deployment scheme for the required SCs to maximize investment efficiency while maintaining the SCR within a desirable range. As an alternative, improving dynamic control strategies can also benefit power systems. For instance, \cite{khan2024grid} proposes switching the control mode of IBR from GFL to GFM to proactively enhance voltage stability. Nevertheless, such a strategy requires carefully coordinated mode transitions; otherwise, abrupt switching may disrupt the control loops of power converters and even jeopardize system stability \cite{kim2025seamless}. 

To avoid new investment costs and control mode switching, \cite{chu2022voltage} develops explicit operating-point constraints for generating units to ensure that the system operates at the required SCR level while maintaining static voltage stability. This method effectively captures the distinct voltage stability support mechanisms of different generator types and fully exploits their respective capabilities. Specifically, synchronous machines, including Synchronous Generators (SGs) and Virtual Synchronous Generators (VSGs), reduce the equivalent network impedance by remaining online, whereas GFL-IBR provide reactive power support when required.

Going further, in order to incentivize generating units to provide such voltage stability support, a market mechanism in \cite{wang2026shadow} has been developed to enable participating units to receive remuneration for providing voltage stability as an ancillary service. Both the `dispatchable' and `restricted' pricing methods (the first consistingof relaxing integer variables, while the second applies a two-stage approach) were evaluated, in order to compute service prices while addressing the challenge of deriving dual variables in the mixed-integer Unit Commitment (UC) problem.

However, both pricing approaches exhibit limitations in addressing integrality while ensuring adequate revenues for units. Dispatchable pricing relaxes binary commitment variables to obtain a convex formulation from which shadow prices can be derived. As a result, thermal units can be `partially committed', causing the resulting prices to reflect only part of their contributions to voltage stability. The insufficient remuneration may therefore lead to negative net profits, potentially driving them out of the market. In contrast, restricted pricing preserves binary commitment decisions and solves the UC twice to derive `commitment prices', which can offset a part of uplift payments that compensate for the fixed cost of SGs. Nevertheless, uplift payments may still be required if energy and ancillary service prices cannot form adequate revenues for SGs.

More importantly, uplift payments are allocated on a unit-specific basis and remain unpublicized in real time \cite{hohl2023intraday}, undermining market transparency. Besides, it remains unclear who should bear these payments and how the associated costs should be distributed, both of which continue to attract research attention. %\cite{lin2023non}.

To overcome these limitations, this paper proposes a novel pricing method for static voltage stability based on the Primal-Dual (P-D) formulation. The framework incorporates explicit profitability constraints accounting for energy and ancillary service prices to ensure non-negative profits for participating units, while recovering binary commitment decisions through a duality-gap minimization process. Consequently, it produces adequate incentives for generators to stay in the market and eliminates the need for uplift payments.

The contributions of this work are:
\begin{enumerate}
    \item To propose a novel pricing method based on the primal-dual formulation for computing shadow prices of static voltage stability services, including prices for SCR enhancement and reactive power support from GFL-IBR.
    \item To establish profit calculation formulations applicable to SGs and IBR in the proposed energy-voltage stability service market, incorporating both short-term operating costs and long-term investment costs.
    \item To demonstrate that the proposed pricing approach consistently produces incentive-adequate shadow prices across a wide range of system conditions, whereas traditional pricing approaches cannot guarantee this property.
\end{enumerate}

The remainder of this paper is structured as follows: Section-\ref{Review of Existing Schemes for Pricing Voltage Stability Services} reviews existing studies on pricing for voltage stability services. Section-\ref{Pricing Voltage Stability Services} presents the theoretical foundation of the proposed P-D pricing framework and develops its mathematical formulation. Section-\ref{Case Studies} carries out case studies. Section-\ref{Conclusion} concludes this work and prospects future work.

\section{State of the Art in Pricing Static Voltage Stability Services}\label{Review of Existing Schemes for Pricing Voltage Stability Services}
This section first presents the static voltage stability constraint for GFL-IBR buses proposed in \cite{chu2022voltage}. It then derives shadow prices for this ancillary service, reviews existing methods for computing these prices \cite{wang2026shadow}, and points out their limitations in addressing non-convexity.

\subsection{Representation of Static Voltage Stability Constraints}\label{Representation of Static Voltage Stability Constraint}

Under steady-state conditions, the voltage stability constraint for GFL-IBR buses reads: %shown in Fig.~\ref{Z_define}
\begin{equation}\label{eq:explanation_voltage_stability}
  P_{g_f}^{2}+Q_{g_f}^{2} \leq \big( Q_{g_f} + \frac{|V^\mathrm{G}_{\Phi(g_f)}|^2}{2|Z_{\Phi(g_f)\Phi(g_f)}|}  \big)^{2}
\end{equation}
where $V^\mathrm{G}_{\Phi(g_f)}$ and $Z_{\Phi(g_f)\Phi(g_f)}$ denote the voltage and the equivalent impedance at the GFL-IBR bus, respectively.

Given that $ |V^\mathrm{G}_{\Phi(g_f)}|^2 \approx 1$ is assumed during normal operation conditions, the constraint \eqref{eq:explanation_voltage_stability} can be further expressed as:

\begin{subequations}\label{eq:define_of_original_VS_constraints}
\begin{align}
    & P_{g_f}^{2}+Q_{g_f}^{2} \leq \big( Q_{g_f} + \frac{1}{2}z_{g_f}  \big)^{2} \\
    & z_{g_f}=\frac{1}{|Z_{\Phi({g_f})\Phi({g_f})}|}
\end{align}
\end{subequations}
where $z_{g_f}$ describes the SCR of bus $\Phi(g_f)$.

The constraints in \eqref{eq:define_of_original_VS_constraints} are in SOC form with respect to $P_{g_f}$, $Q_{g_f}$, and $Q_{g_f}+\frac{1}{2}z_{g_f}$, while the term `$z_{g_f}$' remains highly nonlinear. Specifically, the impedance ratio $1/|Z_{\Phi({g_f})\Phi({g_f})}|$ requires inverting the grid admittance matrix, whose entries depend on the operating states of multiple (V)SGs, such as commitment decisions. Therefore, the offline training approach developed in \cite{chu2022voltage} is employed here to approximately express $z_{g_f}$ in a linear manner, as detailed in Appendix~\ref{VS_cons_approx}.

The adopted formulation shows that voltage stability can be maintained by either adjusting GFL-IBR operating points or strengthening the grid. With the SCR fixed, increased reactive power support from GFL-IBR allows higher active power injection while satisfying voltage constraints. Alternatively, the SCR can be improved, i.e., the equivalent impedance seen by GFL-IBR buses can be reduced by the operation of (V)SGs.

\subsection{Shadow Prices for Static Voltage Stability Services}\label{Assigning Shadow Prices to Static Voltage Stability Services}
To define the associated dual variables, the constraints \eqref{eq:define_of_original_VS_constraints} are first reformulated into the standard SOC form:
\begin{equation}\label{eq:SOC_conver}
    \left\| 
    \begin{bmatrix}
        P_{g_f} \\
        Q_{g_f}
    \end{bmatrix}
    \right\|
    \leq 
        Q_{g_f} +
        \frac{1}{2}z_{{g_f}}: (\lambda_{1,{g_f}}, \lambda_{2,{g_f}}, \mu_{g_f})
\end{equation}
where `$\lambda_{1,{g_f}}$' and `$\lambda_{2,{g_f}}$' correspond to the first and second rows of the left-hand-side matrix in \eqref{eq:SOC_conver}, respectively, and `$\mu_{g_f}$' is associated with the right-hand-side vector. The dual problem is then subject to the following constraint:
\begin{equation}\label{eq:SOC_shadow_price}
\left\|
\begin{aligned}
 \begin{bmatrix}
\lambda_{1,{g_f}} \\
\lambda_{2,{g_f}}
\end{bmatrix}
\end{aligned}
\right\| \leq \mu_{g_f}
\end{equation}
where $\lambda_{1,{g_f}}$ and $\lambda_{2,{g_f}}$ are not bounded.

Accordingly, incorporating the voltage stability constraints \eqref{eq:define_of_original_VS_constraints} into an optimization problem will introduce the following term into the corresponding Lagrangian function:
\begin{equation} \label{eq:SOC_lagrangian}
    \lambda_{1,{g_f}}P_{g_f} + \lambda_{2,{g_f}}Q_{g_f} - \mu_{g_f} (Q_{g_f} + \frac{1}{2}z_{{g_f}} )
\end{equation}

Finally, applying the KKT stationarity conditions to \eqref{eq:SOC_lagrangian} yields the following gradient expression, from which the shadow prices of static voltage stability services at GFL-IBR buses can be derived:
\begin{itemize}
    \item Price of $Q_{g_f}$: \quad  $-\lambda_{2,{g_f}}+\mu_{g_f}$
    \item Price of $\mathrm{SCR}_{\Phi(g_f)}$ enhancement: \quad  $\mu_{g_f}$
\end{itemize}
Note that `$P_{g_f}$' is not remunerated as a voltage stability service, as it primarily satisfies energy demand and instead consumes the available voltage stability margin, thereby potentially degrading voltage security, as discussed in \cite{wang2026shadow}.

\subsection{Existing Methods for Computing Shadow Prices}\label{Existing Methods for Computing Shadow Prices}
This subsection reviews two state-of-the-art pricing methods for voltage stability services and highlights their limitations relative to the proposed P-D formulation.

\subsubsection{Restricted Pricing Method}\label{Restricted Pricing}
Restricted pricing follows a two-stage solution procedure. The original optimization model is first solved to determine the optimal commitment schedule, denoted by `$u_{g_c}^*$'. The model is then re-solved with the integrality constraints relaxed while fixing the commitment variables at their optimal values:
\begin{subequations}\label{eq:restricted_define}
    \begin{align}
   & u_{g_c}=u_{g_c}^*:(\lambda_{g_c,\mathrm{commit}}),~\forall g_c  \\
   & \eta_m=\eta_m^*,~\forall m
   \end{align}
\end{subequations}  
Accordingly, a new term $\lambda_{g_c,\mathrm{commit}}\cdot(u_{g_c}^* - u_{g_c})$ is incorporated into the Lagrangian function, which generates corresponding remuneration for the relevant SGs:
\begin{equation}\label{eq:commitment_price}
    \lambda_{g_c,\mathrm{commit}} \cdot u_{g_c}^*
\end{equation}
where `$\lambda_{g_c,\mathrm{commit}}$' denotes the commitment price, which constitutes only part of the uplift payments associated with committed SGs. Such payments help compensate thermal units for fixed costs that cannot be covered through energy and ancillary service revenues, thereby preserving their incentives to follow dispatch. Nevertheless, how these uplift payments should be allocated among market participants remains an open issue. In the absence of a well-defined cost-allocation mechanism, some participants may bear an unfair share of the uplift charges, potentially reducing market efficiency.

More critically, weather-driven IBR, which have no commitment variables, are ineligible for such compensation, potentially resulting in inadequate profits.

\subsubsection{Dispatchable Pricing Method}\label{Dispatchable Pricing Method}
This method works by relaxing all binary variables to continuous ones in order to calculate shadow prices:
\begin{equation}
   u_{g_c}\in\{0, 1\} \Rightarrow 0 \leq u_{g_c} \leq 1, ~ \forall g_c
\end{equation}
The nonlinear term `$\eta_m$' in \eqref{eq:approx_z_coefficients} consequently becomes a bilinear product of two continuous variables, making the problem non-convex. Although McCormick envelopes can be employed to convexify these bilinear terms, the resulting relaxations are often loose and may enlarge the feasible region. More importantly, relaxing binary commitment variables distorts real-world system operating conditions and may underestimate generators’ contributions to voltage stability, particularly the SCR support provided by SGs, which can only be accurately captured when their commitments are complete. As a result, make-whole payments may be required to compensate for the resulting revenue shortfalls.
        
To address the shortcomings in these two pricing approaches, this paper proposes a method based on the P-D formulation. It aims to preserve the binary variables and derive shadow prices that adequately reward generators without relying on uplift payments. A comparative overview of the various pricing methods is given in Table~\ref{table:pricing-schemes}.

\section{Primal-Dual Formulation for Pricing Static Voltage Stability Services}\label{Pricing Voltage Stability Services}
This section first reviews the mathematical foundation of the P-D formulation and then presents the proposed pricing model based on a voltage stability constrained UC formulated as a Mixed-Integer Second-Order Cone Program (MISOCP).

\begin{table}[t]
\centering
\caption{Main Features of Shadow Pricing Schemes for Static Voltage Stability Services}
\setlength{\tabcolsep}{2.7pt}
{
\fontsize{7.5pt}{10pt}\selectfont % <--- 设置字号为 8pt，行距为 10pt
\renewcommand{\arraystretch}{1.2}
\begin{tabular}{lccc}
\toprule
Pricing schemes & UC property & Cost recovery & Associated dual variables \\
\midrule
Dispatchable    & Relaxed & Conditional & $\lambda_{2,{g_f}},\mu_{g_f}$ \\
Restricted     & Integer  & Conditional   & $\lambda_{2,{g_f}},\mu_{g_f},\lambda_{g_c,\mathrm{commit}}$ \\
P-D formulation  & Integer  & Guaranteed   & $\lambda_{2,{g_f}},\mu_{g_f}$ \\
\bottomrule
\end{tabular}
}
\label{table:pricing-schemes}
\end{table}

\subsection{Mathematical Background of Primal-Dual Formulation}\label{P-D basis}
For the left-hand linear program in \eqref{eq:primal_dual_final}, its dual problem can be derived in the right-hand side.
\begin{equation}\label{eq:primal_dual_final}
\setlength{\arraycolsep}{5pt}
\renewcommand{\arraystretch}{1.0} 
\begin{array}{@{}llc!{\vline}llc@{}}
\min\limits_{x} && c^\mathrm{T} x 
&
\max\limits_{\mu} && b^\mathrm{T} \mu \\
\text{s.t.}      && Ax \geq b,\, x \geq 0 
&
\text{s.t.}      && A^\mathrm{T} \mu \leq c,\, \mu \geq 0
\end{array}
\end{equation}
where $x \in \mathbb{R}^n$, $c \in \mathbb{R}^n$, $A \in \mathbb{R}^{m \times n}$, $b \in \mathbb{R}^m$ and $\mu \in \mathbb{R}^m$, with the shadow prices contained in the dual variables $\mu$. 

According to \cite{ruiz2012pricing}, relaxing the complementary slackness conditions can yield an optimization problem equivalent to \eqref{eq:primal_dual_final}, which is derived as:
\begin{subequations}\label{eq:math_P_D}
\begin{align}
& \min\limits_{x,\mu} \ \epsilon = c^\mathrm{T} x - b^\mathrm{T} \mu
\label{eq:math_P_D_obj}
\\
& \text{s.t.} \quad Ax \geq b,\, x \geq 0; ~ A^\mathrm{T} \mu \leq c,\, \mu \geq 0
\label{eq:math_P_D_con}
\end{align}
\end{subequations}
where \eqref{eq:math_P_D_obj} minimizes the duality gap, subject to the primal and dual constraints in \eqref{eq:math_P_D_con}.

Strong duality is guaranteed for \eqref{eq:primal_dual_final} if $\epsilon=0$. To capture market features and obtain desired solutions, the formulation \eqref{eq:math_P_D} is allowed to accommodate additional constraints, such as binary variables and the linking constraints \eqref{eq:nonnegative_profits_cons} that connect primal variables (e.g., power output) and dual variables (e.g., energy price). Although including these constraints typically compromises strong duality and leads to $\epsilon \geq 0$, the resulting optimal solution remains close to that of \eqref{eq:primal_dual_final}. This is because both primal and dual feasibility are still strictly satisfied and the relaxation is always constructed around the optimal solution for \eqref{eq:primal_dual_final}.

\subsection{Primal of Static Voltage Stability Constrained UC}\label{Primal Voltage Stability Constrained UC Problem}
Without loss of generality, the UC problem formulated as Eq.~\eqref{eq:MISOCP} minimizes system operating costs subject to voltage stability at GFL-IBR buses. Dual variables correspond to the right-hand side of each constraint. As offline training for linear expressions of impedance ratios finishes prior to pricing, $\mathcal{K}$ is not written in \textit{italics} in the P-D formulation.
\begin{subequations}\label{eq:MISOCP}
\begin{align}
& \min_{V_{\mathrm{P}}} \sum_{t} \sum_{g_c} \Big( 
\mathrm{c}_{g_c}^\mathrm{nl} u_{g_c,t} 
+ \mathrm{c}_{g_c}^\mathrm{m} P_{g_c,t} 
+ C_{g_c,t}^\mathrm{st} \Big) 
\label{eq:VS-cons_UC_obj} \\
&  \mathrm{where:} \nonumber \\
&  \quad V_{\mathrm{P}}= \Bigl\{ \quad C_{g_c,t}^\mathrm{st}, u_{g_c,t}, \eta_{m,t}, z_{g_f,t}, \nonumber  \\
                 &  \quad   P_{g_c,t}, P_{g_v,t}, P_{g_f,t},   \nonumber  \\
                 &  \quad   Q_{g_c,t}, Q_{g_v,t}, Q_{g_f,t}   \quad \Bigl\}   \label{eq:LL_var} \\
&  \mathrm{subject~to:} \nonumber \\
%=====================================================
&  C_{g_c,t}^\mathrm{st} \ge 0: (\rho_{g_c,t}^\mathrm{st}),~ \forall g_c, t \label{eq:cons_st_cost_positive} \\
&  C_{g_c,t}^\mathrm{st} \ge (u_{g_c,t}-u_{g_c,t-1})\mathrm{c}^\mathrm{st}_{g_c}:  (\sigma_{g_c,t}^\mathrm{st}),~ \forall g_c, t  \label{eq:cons_st_cost_lb} \\
& u_{g_c,t} \mathrm{P}_{g_c}^\mathrm{min}  \leq P_{g_c,t} \leq u_{g_c,t}  \mathrm{P}_{g_c}^\mathrm{max}: (\tau^{p,\mathrm{lb}}_{g_c,t}, {\tau}^{p,\mathrm{ub}}_{g_c,t}),~  \forall g_c,t \label{eq:cons_SG_output_active} \\ 
& u_{g_c,t} \mathrm{Q}_{g_c}^\mathrm{min}  \leq Q_{g_c,t} \leq u_{g_c,t}  \mathrm{Q}_{g_c}^\mathrm{max}: (\tau^{q,\mathrm{lb}}_{g_c,t}, {\tau}^{q,\mathrm{ub}}_{g_c,t}),~  \forall g_c,t \label{eq:cons_SG_output_reactive} \\ 
& 0 \leq P_{g_v,t} \leq \upalpha_{g_v,t}\mathrm{P}^{\mathrm{max}}_{g_v}: (\tau^{p,\mathrm{lb}}_{g_v,t}, {\tau}^{p,\mathrm{ub}}_{g_v,t}),~  \forall g_v,t \label{eq:cons_VSG_active} \\
&  \mathrm{Q}_{g_v}^\mathrm{min} \leq Q_{g_v,t} \leq \mathrm{Q}_{g_v}^\mathrm{max}: (\tau^{q,\mathrm{lb}}_{g_v,t}, {\tau}^{q,\mathrm{ub}}_{g_v,t}),~  \forall g_v,t \label{eq:cons_VSG_reactive} \\
& 0 \leq P_{g_f,t} \leq \upalpha_{g_f,t}\mathrm{P}^{\mathrm{max}}_{g_f}: (\tau^{p,\mathrm{lb}}_{g_f,t}, {\tau}^{p,\mathrm{ub}}_{g_f,t}),~  \forall g_f,t \label{eq:cons_IBG_active} \\ 
& \mathrm{Q}_{g_f}^\mathrm{min} \leq Q_{g_f,t} \leq \mathrm{Q}_{g_f}^\mathrm{max}: (\tau^{q,\mathrm{lb}}_{g_f,t}, {\tau}^{q,\mathrm{ub}}_{g_f,t}),~  \forall g_f,t \label{eq:cons_IBG_reactive} \\
%==============
%\hspace{-2pt}
%\begin{bmatrix}
%1 & 0 \\
%0 & 1
%\end{bmatrix}
%\hspace{-5pt}
\;&
\left\|
\begin{bmatrix}
P_{g_c,t} \\
Q_{g_c,t}
\end{bmatrix}
\right\| 
\le 
\mathrm{S}_{g_c}^\mathrm{max}: (\upsilon_{1,g_c,t},\upsilon_{2,g_c,t},\upsilon_{3,g_c,t}),~ \forall g_c,t  \label{eq:cons_SG_capacity} \\
%==============
\;&
\left\|
\begin{bmatrix}
P_{g_v,t} \\
Q_{g_v,t}
\end{bmatrix}
\right\|
\le 
\mathrm{S}_{g_v}^\mathrm{max}: (\upsilon_{1,g_v,t}, \hspace{-1pt} \upsilon_{2,g_v,t}, \upsilon_{3,g_v,t}),~ \forall g_v,t  \label{eq:cons_VSG_capacity} \\
%==============
\;&
\left\|
\begin{bmatrix}
P_{g_f,t} \\
Q_{g_f,t}
\end{bmatrix}
\right\|
\le
\mathrm{S}_{g_f}^\mathrm{max}:(\upsilon_{1,g_f,t},\upsilon_{2,g_f,t},\upsilon_{3,g_f,t}),~ \forall g_f,t  \label{eq:cons_IBG_capacity} \\
%==============
 \;& \left\| 
\begin{bmatrix}
P_{g_f,t} \\
Q_{g_f,t}
\end{bmatrix}
\right\|
\hspace{-0.05cm}
\leq 
\hspace{-0.05cm}
Q_{g_f,t} \hspace{-0.05cm} + \hspace{-0.05cm} \frac{1}{2}z_{{g_f,t}}\hspace{-0.1cm}:\hspace{-0.1cm}( \lambda_{1,g_f,t}, \lambda_{2,g_f,t}, \mu_{g_f,t}),~\forall g_f,t  \label{eq:cons_VS_LL} \\
& z_{g_f,t} = \hspace{-0.1cm}
\sum_{g_c \in \mathcal{G}_c}  \hspace{-0.1cm} \mathrm{k}_{{g_f},g_c} u_{g_c,t}
+  \hspace{-0.1cm} \sum_{g_v \in \mathcal{G}_v} \mathrm{k}_{{g_f},g_v} \upalpha_{g_v,t}  \nonumber \\
&   \hspace{0.65cm} +  \sum_{m \in \mathcal{M}} \mathrm{k}_{{g_f},m} \eta_{m,t}: (\xi_{g_f,t}),~ \forall g_f,t \label{eq:define_z} \\
%==============
&  \sum_{i \in \{\Phi(g_c), \Phi(g_v), \Phi(g_f)\}} \hspace{-1cm} P_{i,t} = \sum_i \mathrm{P}_{i,t}^\mathrm{D}: (\lambda^{\mathrm{E}}_t),~ \forall t \label{eq:cons_balance_active} \\
%==============
&  \sum_{i \in \{\Phi(g_c), \Phi(g_v), \Phi(g_f)\}} \hspace{-1cm} Q_{i,t} = \sum_i \mathrm{Q}_{i,t}^\mathrm{D}: (\phi_t),~\forall t \label{eq:cons_balance_reactive} \\
%==============
% &  P_{ij,t} + P_{ji,t} = 0: (\chi_{ij,t}^p),~ \forall i,j \in l(i),t   \label{eq:cons_AC_active_direction} \\
% &  Q_{ij,t} + Q_{ji,t} = 0: (\chi_{ij,t}^q),~ \forall i,j \in l(i),t   \label{eq:cons_AC_reactive_direction} \\
& \eta_{m,t}\leq u_{g_c',t}: (\gamma_{1,m,t}^{\mathrm{max}}),~\forall m\in\{ g_c', g_c''  \}, t \label{eq:MC_linear_1} \\
& \eta_{m,t}\leq u_{g_c'',t}: (\gamma_{2,m,t}^{\mathrm{max}}),~\forall m\in\{ g_c', g_c''  \}, t \label{eq:MC_linear_2} \\
& \eta_{m,t} \geq u_{g_c',t}+u_{g_c'',t}-1: (\gamma_{1,m,t}^{\mathrm{min}}),\forall m\in\{ g_c', g_c''  \}, t \label{eq:MC_linear_3} \\
& \eta_{m,t} = u_{g_c,t}\upalpha_{g_v,t} : (\gamma_{m,t}),~\forall m\in\{ g_c, g_v  \}, t \label{eq:product_UC_vsg_state} \\
& u_{g_c,t} \in \{0,1\},~ \forall g_c,t \label{eq:binary_SGs} \\
& \eta_{m,t} \in \{0, 1\},~\forall m\in\{ g_c', g_c''  \}, t \label{eq:MC_linear_4}
\end{align}  
\end{subequations}   
where the objective function \eqref{eq:VS-cons_UC_obj} accounts for the no-load, generation, and startup costs of SGs. Eq.~\eqref{eq:LL_var} defines the primal variables that follow constraints: Eqs.~\eqref{eq:cons_st_cost_positive}–\eqref{eq:cons_st_cost_lb} define startup costs; Eqs.~\eqref{eq:cons_SG_output_active}–\eqref{eq:cons_SG_output_reactive}, \eqref{eq:cons_VSG_active}–\eqref{eq:cons_VSG_reactive}, and \eqref{eq:cons_IBG_active}–\eqref{eq:cons_IBG_reactive} bound the active and reactive power outputs of SGs, VSGs, and GFL-IBR, respectively; Eqs.~\eqref{eq:cons_SG_capacity}–\eqref{eq:cons_IBG_capacity} constrain generator outputs within their apparent power capacities; Eq.~\eqref{eq:cons_VS_LL} enforces static voltage stability at GFL-IBR buses; Eq.~\eqref{eq:define_z} is the linear expression of the impedance ratio; Eqs.~\eqref{eq:cons_balance_active}–\eqref{eq:cons_balance_reactive} ensure active and reactive power balance; Eqs.~\eqref{eq:MC_linear_1}–\eqref{eq:MC_linear_3} provide auxiliary constraints for linearizing $\eta_{m,t}$ when it represents the product of two commitment variables through McCormick envelopes; Eq.~\eqref{eq:product_UC_vsg_state} defines the product of the operating states of SGs and VSGs; Eqs.~\eqref{eq:binary_SGs}–\eqref{eq:MC_linear_4} impose binary-variable requirements.

\subsection{Dual of Relaxed Static Voltage Stability Constrained UC}\label{Dual of Relaxed Static Voltage Stability Constrained UC}
Since MISOCP \eqref{eq:MISOCP} is non-convex, its dual problem can be derived after relaxing the binary variables:
\begin{subequations}  \label{eq:integer_relax}
\begin{align}
& 0 \leq u_{g_c,t} \leq 1: (\psi_{g_c,t}^{\mathrm{min}}, \psi_{g_c,t}^{\mathrm{max}}),~\forall g_c,t \\
& \eta_{m,t} \geq 0: (\gamma_{2,m,t}^{\mathrm{min}}),~\forall m, t \label{eq:MC_linear_5}
\end{align}
\end{subequations}
where the upper bound of $\eta_{m,t}$ is determined by $u_{g_c',t}$, $u_{g_c'',t}$ or $\upalpha_{g_v,t}$, as indicated by \eqref{eq:MC_linear_1}, \eqref{eq:MC_linear_2} and \eqref{eq:product_UC_vsg_state}.

The dual formulation of relaxed problem \eqref{eq:MISOCP} can then be derived as follows:
\begin{subequations}\label{eq:dual_MISOCP}
\begin{align}
&  \max_{V_{\mathrm{D}}}    \sum_{t} \Big[ -\hspace{-0.2cm}\sum_{g_c\in\mathcal{G}_c} \hspace{-0.2cm} \big( \mathrm{S}_{g_c}^{\mathrm{max}}\upsilon_{3,g_c,t} + \psi_{g_c,t}^{\mathrm{max}} \big)- \hspace{-0.2cm}\sum_{g_v\in\mathcal{G}_v} \hspace{-0.2cm}\big(\upalpha_{g_v,t}\mathrm{P}^{\mathrm{max}}_{g_v}\tau^{p,\mathrm{ub}}_{g_v,t} \nonumber \\ 
& + \mathrm{Q}_{g_v}^{\mathrm{max}}\tau^{q,\mathrm{ub}}_{g_v,t} - \mathrm{Q}_{g_v}^{\mathrm{min}}\tau^{q,\mathrm{lb}}_{g_v,t} + \mathrm{S}_{g_v}^{\mathrm{max}}\upsilon_{3,g_v,t} \big) - \hspace{-0.2cm} \sum_{g_f\in\mathcal{G}_f} \hspace{-0.2cm} \big( \upalpha_{g_f,t}\mathrm{P}^{\mathrm{max}}_{g_f}\tau^{p,\mathrm{ub}}_{g_f,t} \nonumber \\ 
& + \mathrm{Q}_{g_f}^{\mathrm{max}}\tau^{q,\mathrm{ub}}_{g_f,t} - \mathrm{Q}_{g_f}^{\mathrm{min}}\tau^{q,\mathrm{lb}}_{g_f,t} + \mathrm{S}_{g_f}^{\mathrm{max}}\upsilon_{3,g_f,t} - \hspace{-0.2cm} \sum_{g_v\in\mathcal{G}_v}\hspace{-0.15cm}\mathrm{k}_{g_f,g_v}\upalpha_{g_v,t}\xi_{g_f,t} \big)  \nonumber \\
& + \sum_i \big(\mathrm{P}^{\mathrm{D}}_{i,t} \lambda^{\mathrm{E}}_{t}  +  \mathrm{Q}^{\mathrm{D}}_{i,t} \phi_{t} \big) - \hspace{-0.1cm}\sum_m\gamma^{\mathrm{min}}_{1,m,t} \Big] \hspace{-0.1cm} - \hspace{-0.2cm} \sum_{g_c\in\mathcal{G}_c} \hspace{-0.1cm}u_{g_c,0}\mathrm{c}^{\mathrm{st}}_{g_c}\sigma_{g_c,t=1}^\mathrm{st}  \label{eq:DLL_obj}   \\
&  \mathrm{where:} \nonumber \\
%===============================================
& V_\mathrm{D}  = \{ \quad \sigma_{g_c,t}^\mathrm{st}, \tau^{\{ \cdot \}}_{g_c,t}, \tau^{p,\mathrm{ub}}_{g_v,t}, \tau^{q,\mathrm{ub}}_{g_v,t}, \tau^{q,\mathrm{lb}}_{g_v,t}, \tau^{p,\mathrm{ub}}_{g_f,t}, \tau^{q,\mathrm{ub}}_{g_f,t}, \tau^{q,\mathrm{lb}}_{g_f,t}, \nonumber \\ 
&    \upsilon_{ \{ \cdot \},t }, \lambda_{ \{ \cdot \},g_f,t }, \mu_{g_f,t}, \xi_{g_f,t}, \lambda^{\mathrm{E}}_t, \phi_t, \nonumber \\
&   \gamma_{1,m,t}^{\mathrm{max}}, \gamma_{2,m,t}^{\mathrm{max}}, \gamma_{1,m,t}^{\mathrm{min}}, \gamma_{m,t}, \psi_{g_c,t}^{\mathrm{max}} \quad \} 
\label{eq:DLL_variables} 
\end{align} 
\begin{align}
&  \mathrm{subject~to:} \nonumber \\
& 1 - \sigma_{g_c,t}^\mathrm{st} \ge 0,~ \forall g_c, t \label{eq:DLL_cons_star_cost} \\
%==============
& \mathrm{c}_{g_c}^\mathrm{nl} + \psi_{g_c,t}^{\mathrm{max}} - \mathrm{P}_{g_c}^{\mathrm{max}}\tau^{p,\mathrm{ub}}_{g_c,t} + \mathrm{P}_{g_c}^{\mathrm{min}} \tau^{p,\mathrm{lb}}_{g_c,t} - \mathrm{Q}_{g_c}^{\mathrm{max}}\tau^{q,\mathrm{ub}}_{g_c,t} \nonumber \\  
& + \mathrm{Q}_{g_c}^{\mathrm{min}} \tau^{q,\mathrm{lb}}_{g_c,t} + \sum_{g_f \in \mathcal{G}_f} \mathrm{k}_{g_f,g_c} \xi_{g_f,t} + \hspace{-0.25cm} \sum_{m \in \{g_c,g_v\} } \hspace{-0.25cm} \upalpha_{g_v,t}\gamma_{m,t}  \nonumber \\ 
&  +  \mathrm{c}_{g_c}^\mathrm{st} \sigma_{g_c,t}^\mathrm{st} + h_{g_c,t}  \ge 0,   ~ \forall g_c, t = T   \label{eq:DLL_cons_binary_stra_commit_SG_t1} \\
%==============
& \mathrm{c}_{g_c}^\mathrm{nl} + \psi_{g_c,t}^{\mathrm{max}} - \mathrm{P}_{g_c}^{\mathrm{max}}\tau^{p,\mathrm{ub}}_{g_c,t} + \mathrm{P}_{g_c}^{\mathrm{min}} \tau^{p,\mathrm{lb}}_{g_c,t} - \mathrm{Q}_{g_c}^{\mathrm{max}}\tau^{q,\mathrm{ub}}_{g_c,t} \nonumber \\  
& + \mathrm{Q}_{g_c}^{\mathrm{min}} \tau^{q,\mathrm{lb}}_{g_c,t} + \sum_{g_f \in \mathcal{G}_f} \mathrm{k}_{g_f,g_c} \xi_{g_f,t} + \hspace{-0.25cm} \sum_{m \in \{g_c,g_v\} } \hspace{-0.25cm} \upalpha_{g_v,t}\gamma_{m,t} \nonumber \\
&  - \mathrm{c}_{g_c}^\mathrm{st} \sigma_{g_c,t+1}^\mathrm{st} + \mathrm{c}_{g_c}^\mathrm{st} \sigma_{g_c,t}^\mathrm{st} + h_{g_c,t}  \ge 0,   ~ \forall g_c, t \leq T-1   \label{eq:DLL_cons_binary_stra_commit_SG_t_2_T} \\
%============== 
& \sum_{g_f \in \mathcal{G}_f} \mathrm{k}_{{g_f},m} \xi_{g_f,t} + \gamma_{1,m,t}^{\mathrm{max}} + \gamma_{2,m,t}^{\mathrm{max}} \nonumber \\
& - \gamma_{1,m,t}^{\mathrm{min}} \ge 0,~ \forall m\in\{ g_c', g_c''  \}, t \label{eq:DLL_cons_eta_m_SGs} \\
%============== 
& \sum_{g_f \in \mathcal{G}_f} \mathrm{k}_{{g_f},m} \xi_{g_f,t} - \gamma_{m,t} = 0,~ \forall m\in\{ g_c, g_v \}, t \label{eq:DLL_cons_eta_m_SG_VSG} \\
%================ 
&  -\frac{1}{2}\mu_{g_f,t} - \xi_{g_f,t} = 0,~ \forall g_f, t   \label{eq:DLL_cons_z} \\
%================ 
&  \mathrm{c}_{g_c}^\mathrm{m} + \tau^{p,\mathrm{ub}}_{g_c,t} - \tau^{p,\mathrm{lb}}_{g_c,t} - \upsilon_{1,g_c,t} - \lambda^{\mathrm{E}}_t = 0,~ \forall g_c, t   \label{eq:DLL_cons_P_strategic_SG} \\
%================ 
&  \tau^{p,\mathrm{ub}}_{g_v,t} - \upsilon_{1,g_v,t} - \lambda^{\mathrm{E}}_t = 0,~ \forall g_v, t   \label{eq:DLL_cons_P_VSG} \\
%================ 
&  \tau^{p,\mathrm{ub}}_{g_f,t} - \upsilon_{1,g_f,t} - \lambda_{1,g_f,t} - \lambda^{\mathrm{E}}_t  = 0,~ \forall g_f, t   \label{eq:DLL_cons_P_strategic_IBG} \\
%================ 
&  \tau^{q,\mathrm{ub}}_{g_c,t} - \tau^{q,\mathrm{lb}}_{g_c,t} - \upsilon_{2,g_c,t} - \phi_t = 0,~ \forall g_c, t   \label{eq:DLL_cons_Q_strategic_SG} \\
%================ 
&  \tau^{q,\mathrm{ub}}_{g_v,t} - \tau^{q,\mathrm{lb}}_{g_v,t} - \upsilon_{2,g_v,t} - \phi_{t} = 0,~ \forall g_v, t   \label{eq:DLL_cons_Q_VSG} \\
%================ 
&  \tau^{q,\mathrm{ub}}_{g_f,t} - \tau^{q,\mathrm{lb}}_{g_f,t} \hspace{-0.09cm}  - \hspace{-0.09cm} \upsilon_{2,g_f,t} \hspace{-0.09cm}  - \hspace{-0.09cm} \lambda_{2,g_f,t} \hspace{-0.09cm}  - \hspace{-0.09cm} \mu_{g_f,t} \hspace{-0.09cm}  - \hspace{-0.09cm} \phi_{t}  = 0,~ \forall g_f, t   \label{eq:DLL_cons_Q_strategic_IBG} \\ 
%============== 
            &
\left\|
\begin{bmatrix}
\upsilon_{1,g_c,t} \\
\upsilon_{2,g_c,t}
\end{bmatrix}
\right\|
\le
\upsilon_{3,g_c,t},~ \forall g_c,t  \label{eq:dual_cons_SG_capacity} \\
%================
        &
\left\|
\begin{bmatrix}
\upsilon_{1,g_v,t} \\
\upsilon_{2,g_v,t}
\end{bmatrix}
\right\|
\le
\upsilon_{3,g_v,t},~ \forall g_v,t  \label{eq:dual_cons_VSG_capacity} \\
%================
    &
\left\|
\begin{bmatrix}
\upsilon_{1,g_f,t} \\
\upsilon_{2,g_f,t}
\end{bmatrix}
\right\|
\le
\upsilon_{3,g_f,t},~ \forall g_f,t  \label{eq:dual_cons_IBG_capacity} \\
%================
    &
\left\|
\begin{bmatrix}
\lambda_{1,{g_f},t} \\
\lambda_{2,{g_f},t}
\end{bmatrix}
\right\|
\le  
\mu_{{g_f},t},~ \forall g_f,t  \label{eq:dual_cons_VSS} \\
%================
    & \{V_\mathrm{D} | V_\mathrm{D} \neq \xi_{g_f,t}, \lambda_{t}^{\mathrm{E}}, \phi_{t}, \gamma_{m,t}, \lambda_{\{\cdot\},{g_f},t}, \upsilon_{\{ \cdot \},t} \}   \in \mathbb{R}_+ \label{eq:DLL_var_nonnegative}
\end{align}
\end{subequations}  
where \eqref{eq:DLL_obj} is the objective function of the dual. Eq.~\eqref{eq:DLL_variables} lists the involved dual variables. The correspondence between dual constraints and primal variables is shown as follows: Eq.~\eqref{eq:DLL_cons_star_cost}$\leftrightarrow$$C_{g_c,t}^\mathrm{st}$; Eqs.~\eqref{eq:DLL_cons_binary_stra_commit_SG_t1}-\eqref{eq:DLL_cons_binary_stra_commit_SG_t_2_T}$\leftrightarrow$$u_{g_c,t}$; Eqs.~\eqref{eq:DLL_cons_eta_m_SGs}-\eqref{eq:DLL_cons_eta_m_SG_VSG}$\leftrightarrow$$\eta_{m,t}$; Eq.~\eqref{eq:DLL_cons_z}$\leftrightarrow$$z_{g_f,t}$; Eq.~\eqref{eq:DLL_cons_P_strategic_SG}$\leftrightarrow$$P_{g_c,t}$; Eq.~\eqref{eq:DLL_cons_P_VSG}$\leftrightarrow$$P_{g_v,t}$; Eq.~\eqref{eq:DLL_cons_P_strategic_IBG}$\leftrightarrow$$P_{g_f,t}$; Eq.~\eqref{eq:DLL_cons_Q_strategic_SG}$\leftrightarrow$$Q_{g_c,t}$;  Eq.~\eqref{eq:DLL_cons_Q_VSG}$\leftrightarrow$$Q_{g_v,t}$; \\ Eq.~\eqref{eq:DLL_cons_Q_strategic_IBG}$\leftrightarrow$$Q_{g_f,t}$; Eqs.~\eqref{eq:dual_cons_SG_capacity}–\eqref{eq:dual_cons_VSS} define the dual variables corresponding to the primal SOC constraints within the cone; Eq.~\eqref{eq:DLL_var_nonnegative} states the dual variables that are not confined to be non-negative. The term $h_{g_c,t}$ in \eqref{eq:DLL_cons_binary_stra_commit_SG_t1}-\eqref{eq:DLL_cons_binary_stra_commit_SG_t_2_T} represents the dual parts associated with auxiliary equations in \eqref{eq:MC_linear_1}-\eqref{eq:MC_linear_3}, which are detailed in Appendix~\ref{Dual constraints for UC states in McCormick envelopes}.

\subsection{Primal-Dual Formulation for Pricing Static Voltage Stability Services}\label{Primal-Dual Formulation}
The proposed P-D formulation seeks to derive revenue-adequate prices `$\lambda_t^\mathrm{E}$', `$\mu_{g_f,t}$', and `$-\lambda_{2,{g_f},t}+\mu_{g_f,t}$' under nonconvex operating conditions. These prices are intended to provide sufficient economic incentives for generation units and ensure that each unit can fully recover its costs through the provision of both energy and ancillary services. Accordingly, as discussed in Section~\ref{P-D basis}, non-negative profit constraints below are incorporated into the pricing:
\begin{subequations}\label{eq:nonnegative_profits_cons}
\begin{align}
& \sum_t \Big(\underbrace{\lambda_t^\mathrm{E} P_{g_c,t}}_{\text{SG's energy revenue}} + \quad \underbrace{\lambda_{g_c,\mathrm{commit},t} \cdot u_{g_c,t}^* }_{\text{Commitment revenue}} \hspace{-0.2cm} \nonumber \\
& + \hspace{-0.1cm} \underbrace{\sum_{g_f \in \mathcal{G}_f} \hspace{-0.2cm} \mu_{g_f,t} \mathrm{k}_{g_f,g_c} u_{g_c,t}
    + \hspace{-0.1cm} \sum_{g_f \in \mathcal{G}_f} \sum_{\{m \mid g_c \in m \}} \hspace{-0.4cm} \mu_{g_f,t} \mathrm{k}_{g_f,m} \eta_{m,t}}_{\text{SG's voltage stability service revenue}} \nonumber \\ 
& - \underbrace{\mathrm{c}_{g_c}^\mathrm{nl} u_{g_c,t} + \mathrm{c}_{g_c}^\mathrm{m} P_{g_c,t}
    + C_{g_c,t}^\mathrm{st}}_{\text{SG's operating cost}} \Big) \geq 0, \quad \forall g_c \label{eq:nonnegative_profits_cons_SGs} \\ 
& \sum_t \Big(\underbrace{\lambda_t^\mathrm{E} P_{g_v,t}}_{\text{VSG's energy revenue}} \hspace{-0.2cm} \nonumber \\ 
& + \hspace{-0.1cm} \underbrace{ \sum_{g_f \in \mathcal{G}_f} \hspace{-0.2cm} \mu_{g_f,t} \mathrm{k}_{g_f,g_v} \upalpha_{g_v,t}
    + \hspace{-0.1cm} \sum_{g_f \in \mathcal{G}_f} \sum_{\{m \mid g_v \in m \}} \hspace{-0.4cm} \mu_{g_f,t} \mathrm{k}_{g_f,m} \eta_{m,t}  }_{\text{VSG's voltage stability service revenue}} \Big) \nonumber \\ 
&    - \mathrm{LCOE}_{g_v} \geq 0, \quad \forall g_v \label{eq:nonnegative_profits_cons_VSG} \\
& \hspace{-1cm} \sum_t \Big(\underbrace{\lambda_t^\mathrm{E} P_{g_f,t}}_{\text{GFL-IBR's energy revenue}} \hspace{-0.2cm} + \underbrace{ (-\lambda_{2,{g_f,t}} + \mu_{g_f,t})Q_{g_f,t}  }_{\text{GFL-IBR's voltage stability service revenue}} \Big) \nonumber \\ 
& - \mathrm{LCOE}_{g_f} \geq 0, \quad \forall g_f \label{eq:nonnegative_profits_cons_IBR}
\end{align}
\end{subequations}
where the commitment price remuneration in \eqref{eq:nonnegative_profits_cons_SGs} is applied only in the restricted pricing. The Levelized Cost of Energy (LCOE) in \eqref{eq:nonnegative_profits_cons_VSG} and \eqref{eq:nonnegative_profits_cons_IBR} quantifies daily amortized capital expenditures for VSGs and GFL-IBR. This upfront capital investment is evenly distributed across the asset lifespan on a daily basis, resulting in a daily cost (in €/day) independent of wind power generation \cite{ho2021regional}.

By introducing LCOE into pricing, the capital investment cost of IBR is captured, overcoming the limitation of pricing based solely on short-run operating costs. The combined revenues from energy and ancillary services cover the full lifecycle cost, ensuring revenue adequacy and incentivizing investment in IBR at system locations in need.

Finally, the P-D formulation for optimizing market clearing prices is compactly expressed as follows:
\begin{subequations}\label{eq:final_pricing}
 \begingroup
\begin{alignat}{2}
&  \displaystyle \min_{V}~\eqref{eq:VS-cons_UC_obj}-\eqref{eq:DLL_obj}   \label{eq:final_pricing_obj} 
\end{alignat}
\endgroup
where:   
\begin{alignat}{3}
&   V= \Bigl\{~V_\mathrm{P}~\eqref{eq:LL_var},~V_\mathrm{D}~\eqref{eq:DLL_variables}~\Bigl\}     \label{eq:final_pricing_variables} 
\end{alignat}
subject to:
\begin{alignat}{4}  
    & \mathrm{Primal constraints:}~ \eqref{eq:cons_st_cost_positive}\mathrm{-}\eqref{eq:MC_linear_4} \\ 
    & \mathrm{Dual constraints:}~ \eqref{eq:DLL_cons_star_cost}\mathrm{-}\eqref{eq:DLL_var_nonnegative} \\
    & \mathrm{Non-negative profit constraints:}~\eqref{eq:nonnegative_profits_cons}
\end{alignat}  
\end{subequations}
Note that constraints \eqref{eq:nonnegative_profits_cons} contain nonlinear terms, such as `$\lambda_t^\mathrm{E} P_{g_c,t}$' and `$\mu_{g_f,t} u_{g_c,t}$'. These bilinear expressions introduce additional computational complexity into the formulation. To improve computational tractability, they are linearized prior to solving the model, as described in Appendix~\ref{Linearization of Nonlinear Terms}.

After linearization, the resulting optimization takes the form of an MISOCP, as described in Algorithm~\ref{alg:primal-dual}.

\begin{algorithm}[t]
\caption{Primal-Dual Formulation for Pricing Static Voltage Stability Services}
\label{alg:primal-dual}
\begin{algorithmic}
\setlength{\baselineskip}{1.2\baselineskip}  % 在环境内部开头加这一行
\small  % 整体缩小字体
\Require Static voltage stability-constrained UC: Eqs.~\eqref{eq:MISOCP}
\Ensure  Primal-dual formulation for pricing: Eqs.~\eqref{eq:final_pricing}

%\State \textbf{Step 1:} Approximate system impedance ratio in voltage \\
                        %\hspace{1cm}~ stability constraints via \textit{Eqs. (3)}

\State \textbf{Step 1:} Relax binary variables $u_{g_c,t}$, $\eta_{m,t}$ via Eqs.~\eqref{eq:integer_relax}

\State \textbf{Step 2:} Derive dual of relaxed voltage stability-constrained \\
                        \hspace{1.1cm}UC: Eqs.~\eqref{eq:dual_MISOCP} 

\State \textbf{Step 3:} Set non-negative profit constraints: Eqs.~\eqref{eq:nonnegative_profits_cons}
       
\State \textbf{Step 4:} Merge primal~\eqref{eq:MISOCP} and dual~\eqref{eq:dual_MISOCP}  with  constraints~\eqref{eq:nonnegative_profits_cons} \\
                        \hspace{1cm} to obtain final primal-dual  formulation (Relaxed $u_{g_c,t}$, \\
                        \hspace{1.1cm}$\eta_{m,t}$ are discretized, bilinear terms are linearized)
\end{algorithmic}
\end{algorithm}

\section{Case Studies}\label{Case Studies}
Before analyzing the pricing results, this section first introduces the adopted model parameters and simulation settings.

\subsection{Test System Setting}

 \begin{table}[t]
\centering
\caption{Operation Characteristics of Thermal Units}
\setlength{\tabcolsep}{4pt}
{\fontsize{8pt}{12pt}\selectfont
\begin{tabular}{lcccccc}
\toprule
Bus & 2 & 3 & 4 & 5 & 27 & 30 \\
\midrule   
\(\mathrm{c}_{g_c}^\mathrm{nl}\) (\texteuro/h) & 70 & 60 & 55 & 44 & 40 & 34 \\
\(\mathrm{c}_{g_c}^\mathrm{m}\) (\texteuro/MWh) & 6 & 7 & 11 & 12 & 14 & 15 \\
\(\mathrm{c}_{g_c}^\mathrm{st}\) (\texteuro/h) & 400 & 250 & 185 & 144 & 110 & 62  \\
\(\mathrm{P}_{g_c}^\mathrm{min}\) (MW) & 26 & 23 & 12 & 5 & 5 & 2 \\
\(\mathrm{P}_{g_c}^\mathrm{max}\) (MW) & 53 & 46 & 30 & 27 & 26 & 23 \\
\(u_{g_c,0}\) & 1 & 1 & 0 & 0 & 0  & 0 \\
\bottomrule
\end{tabular}
}
\label{table:units_para}
\end{table}

\begin{table}[t]
\centering
\caption{MAPE of Approximation Errors for Nonlinear Terms in Static Voltage Stability Constraints}
\setlength{\tabcolsep}{6pt} % 可调整列间距
{\fontsize{8pt}{12pt}\selectfont
\begin{tabular}{lcc}
\toprule
Approximated terms & $\frac{1}{|Z_{23,23}|}$ & $\frac{1}{|Z_{24,24}|}$ \\
\midrule
\hspace{0.6cm} MAPE & $1.79$\% & $1.36$\% \\
\bottomrule
\end{tabular}
}
\label{table:Approximation_Errors}
\end{table}

Case studies are conducted on the modified IEEE 30-bus test network referenced in \cite{wang2026shadow} to evaluate the performance of the proposed P-D pricing. Wind generation units are connected to buses \{1, 23, 24\}: the wind unit at bus 1 adopts VSG control, while the other two wind units operate under GFL-IBR modes. Conventional SGs are placed at buses \{2, 3, 4, 5, 27, 30\}. Generating units of different types at each bus follow a specific naming convention; for example, the thermal unit at bus 2 is denoted as `$g_c\mathrm{-}b2$'. Table~\ref{table:units_para} summarizes operational parameters for SGs. The aggregated wind installed capacity reaches 420 MW, and the system base power is defined as $\mathrm{S}_{\mathrm{B}}=\mathrm{100}~\mathrm{MVA}$. The \texttt{Julia-JuMP} implementation developed for this study is publicly available in \cite{Code}. All numerical experiments were conducted on an Apple MacBook Air (M1, 2020) using \texttt{Gurobi 12.0.1} as the MISOCP solver. %The system load varies within the range $\in [\mathrm{176}, \mathrm{451}]$ MW.

For the test system, the offline training process in \eqref{eq:approx_z} is completed in 3.99 s. The Mean Absolute Percentage Error (MAPE) of the approximated terms in the voltage stability constraints is reported in Table~\ref{table:Approximation_Errors}. The overall MAPE remains below the acceptable threshold of 5\% defined in \cite{chu2022voltage}, demonstrating the effectiveness of the approximation.

\subsection{Market Outcomes under Different Pricing Approaches}\label{Market Outcomes under Different Pricing Approaches}

\pgfplotstableread{
G H V
1 1  0.0000
1 2  0.0000
1 3  0.0000
1 4  0.0000
1 5  0.0000
1 6  0.0000
1 7  0.4034
1 8  0.2913
1 9  0.2367
1 10 0.2420
1 11 0.2393
1 12 0.2341
1 13 0.2275
1 14 0.1807
1 15 0.1714
1 16 0.1609
1 17 0.1516
1 18 0.1437
1 19 0.2761
1 20 0.3408
1 21 0.3447
1 22 0.0000
1 23 0.0000
1 24 0.0000
2 1  0.0000
2 2  0.0000
2 3  0.0000
2 4  0.0000
2 5  0.0000
2 6  0.0000
2 7  0.0000
2 8  0.3138
2 9  0.2615
2 10 0.2666
2 11 0.2640
2 12 0.2589
2 13 0.2526
2 14 0.2040
2 15 0.1951
2 16 0.1849
2 17 0.1760
2 18 0.1684
2 19 0.3069
2 20 0.0000
2 21 0.0000
2 22 0.0000
2 23 0.0000
2 24 0.0000
3 1  0.0000
3 2  0.0000
3 3  0.0000
3 4  0.0000
3 5  0.0000
3 6  0.0000
3 7  0.0000
3 8  0.3440
3 9  0.2881
3 10 0.2932
3 11 0.2906
3 12 0.2855
3 13 0.2791
3 14 0.2279
3 15 0.2189
3 16 0.2087
3 17 0.1997
3 18 0.1920
3 19 0.0000
3 20 0.0000
3 21 0.0000
3 22 0.0000
3 23 0.0000
3 24 0.0000
4 1  0.0000
4 2  0.0000
4 3  0.0000
4 4  0.0000
4 5  0.0000
4 6  0.0000
4 7  0.0000
4 8  0.0000
4 9  0.2745
4 10 0.2798
4 11 0.2772
4 12 0.2719
4 13 0.2653
4 14 0.2140
4 15 0.2048
4 16 0.1943
4 17 0.1851
4 18 0.1772
4 19 0.0000
4 20 0.0000
4 21 0.0000
4 22 0.0000
4 23 0.0000
4 24 0.0000
5 1  0.0000
5 2  0.0000
5 3  0.0000
5 4  0.0000
5 5  0.0000
5 6  0.0000
5 7  0.0000
5 8  0.0000
5 9  0.0000
5 10 0.0000
5 11 0.0000
5 12 0.0000
5 13 0.0000
5 14 0.0000
5 15 0.0000
5 16 0.0000
5 17 0.0000
5 18 0.0000
5 19 0.0000
5 20 0.0000
5 21 0.0000
5 22 0.0000
5 23 0.0000
5 24 0.0000
6 1  0.5158
6 2  0.0000
6 3  0.0000
6 4  0.5118
6 5  0.0000
6 6  0.5206
6 7  0.0000
6 8  0.0000
6 9  0.0000
6 10 0.0000
6 11 0.0000
6 12 0.0000
6 13 0.0000
6 14 0.3724
6 15 0.3668
6 16 0.3603
6 17 0.3547
6 18 0.3498
6 19 0.4344
6 20 0.4761
6 21 0.4785
6 22 0.5198
6 23 0.5223
6 24 0.5247
7 1  0.4888
7 2  0.5756
7 3  0.5996
7 4  0.5247
7 5  0.5516
7 6  0.4457
7 7  0.3405
7 8  0.1567
7 9  0.0730
7 10 0.0618
7 11 0.0674
7 12 0.0786
7 13 0.0927
7 14 0.0781
7 15 0.0921
7 16 0.1081
7 17 0.1221
7 18 0.1341
7 19 0.3219
7 20 0.4050
7 21 0.3874
7 22 0.4529
7 23 0.4313
7 24 0.4097
}\dataB
 
\begin{figure}[t]
\centering
\begin{tikzpicture}

\pgfplotsset{
  hmap/.style={
    colormap={bluewhite}{
      color(0)    = (white)
      color(0.20) = (cyan!30!white)
      color(0.45) = (blue!50!white)
      color(0.70) = (blue!80!white)
      color(1)    = (blue!95!black)
    },
    point meta min=0,
    point meta max=0.62,
    width=0.82\columnwidth, 
    height=4.2cm,           
    xlabel={\scriptsize Time (h)},
    ylabel={\scriptsize Units index},
    xlabel style={font=\scriptsize, yshift=2pt},
    ylabel style={font=\scriptsize, xshift=2pt, yshift=-2pt},
    xtick={4,8,12,16,20},
    xmin=1, xmax=24,
    xticklabel style={font=\scriptsize}, 
    ytick={1,2,3,4,5,6,7},
    yticklabels={$g_c\mathrm{-}b2$,$g_c\mathrm{-}b3$,$g_c\mathrm{-}b4$,$g_c\mathrm{-}b5$,$g_c\mathrm{-}b27$,$g_c\mathrm{-}b30$,$g_v\mathrm{-}b1$},
    yticklabel style={font=\scriptsize},
    enlarge x limits={abs=0.5},
    enlarge y limits={abs=0.5},
  }
}

% ---- 核心图表: Node 24 ----
\begin{axis}[
  hmap,
  name=mainPlot,
%  minor xtick={1,2,...,24},      % 次刻度：每 1 小时设一个刻度点
%  grid=both,              % 开启横向和纵向网格线
%  grid style={dashed, gray!50, line width=0.1pt}, % 设置网格线样式（虚线、灰色、细线）
%  axis on top,            % 关键：强制让网格线和边框显示在热力图色块之上
  colorbar=false,
]
\addplot[
  matrix plot*,
  point meta=explicit,
  mesh/cols=24,
  mesh/rows=7,
 ] table[x=H, y=G, meta=V] {\dataB};
\end{axis}

% ---- 颜色条 (高度已单独缩短) ----
\begin{axis}[
  name=cbar,
  % 通过 centered 锚点，让缩短后的颜色条在主图右侧纵向居中
  at={(mainPlot.east)},
  anchor=west,
  xshift=-0.1cm, 
  yshift=-0.8cm, 
  hide axis,
  scale only axis,
  width=0pt,
  height=4.2cm, 
  colormap={bluewhite}{
    color(0)    = (white)
    color(0.20) = (cyan!30!white)
    color(0.45) = (blue!50!white)
    color(0.70) = (blue!80!white)
    color(1)    = (blue!95!black)
  },
  point meta min=0,
  point meta max=0.62,
  colorbar,
  colorbar style={
    width=6pt,          % 宽度保持原样 (6pt)
    height=2.6cm,       % ---- 核心修改：单独缩短颜色条高度 ----
    ytick={0,0.1,0.2,0.3,0.4,0.5,0.6},
    yticklabel style={font=\scriptsize}, 
    ylabel={\scriptsize SCR contributions},
    ylabel style={font=\scriptsize, yshift=3pt}, 
  },
]
\addplot[draw=none] coordinates {(0,0)};
\end{axis}
\end{tikzpicture}
\caption{SCR contributions of $\mathcal{G}_c$ and $\mathcal{G}_v$ to bus 24 over 24 hours.}
\label{fig:scr_node24}
\end{figure}
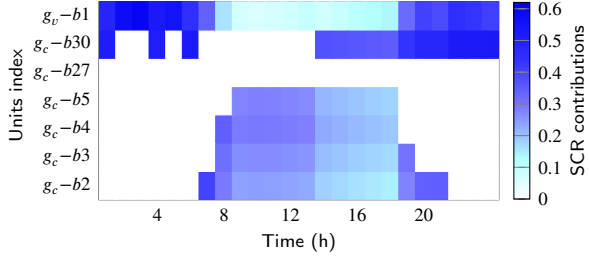

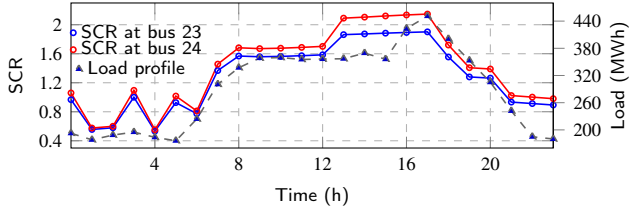
\begin{figure}[t]
\centering
\begin{adjustbox}{minipage=[t]{1\linewidth},center}
\begin{tikzpicture}
% ========== 左Y轴：SCR ==========
\begin{axis}[
    width=0.95\linewidth, height=3.5cm,
    xmin=0, xmax=23,
    ymin=0.3, ymax=2.3,
    xlabel={\scriptsize Time (h)},
    xlabel style={yshift=-8pt},
    ylabel={\scriptsize SCR},
    xtick={4,8,12,16,20},
    ytick={0.4,0.8,1.2,1.6,2.0},
    tick label style={font=\scriptsize},
    axis y line*=left,
    axis x line*=bottom,
    axis lines=box,
    grid=both, grid style={dashed, gray!60},
    legend style={
        font=\scriptsize,
        fill=none,
        draw=none,
        at={(0.3,0.93)}, anchor=north east,
        row sep=-4pt
    },
    legend cell align={left},
    legend image post style={scale=1},
    legend image code/.code={%
        \draw[mark repeat=1,mark phase=1,#1] plot coordinates {(0,0)};
    }
]

% 节点23 SCR（第一条曲线）
\addplot+[color=blue, mark=o, mark size=1pt,line width=0.6pt] coordinates {
(0,0.9669012316325242)
(1,0.556046974928074)
(2,0.5792155988834096)
(3,1.000492628282788)
(4,0.5328783509727345)
(5,0.926591555652211)
(6,0.7729093529492009)
(7,1.368354109241133)
(8,1.5681508042254166)
(9,1.5586503004133898)
(10,1.5634005523194021)
(11,1.5729010561314345)
(12,1.5847766858964674)
(13,1.8632990001234064)
(14,1.8728927145423924)
(15,1.8838569595926427)
(16,1.8934506740116248)
(17,1.9016738577993078)
(18,1.5553526500988555)
(19,1.2801849034609962)
(20,1.2640752459612912)
(21,0.933309834982262)
(22,0.913154996992105)
(23,0.8930001590019485)
};
\addlegendentry{SCR at bus 23}

% 节点24 SCR（第二条曲线）
\addplot+[color=red, mark=o, mark size=1pt, line width=0.6pt] coordinates {
(0,1.0594151500337545)
(1,0.575588165442074)
(2,0.5995710056688249)
(3,1.0953568154494842)
(4,0.551605325215319)
(5,1.0162851515348774)
(6,0.8111070461429406)
(7,1.4568575240778985)
(8,1.6819232815390215)
(9,1.670691130443612)
(10,1.6763072059913198)
(11,1.6875393570867332)
(12,1.701579545955998)
(13,2.0908220363380354)
(14,2.1048366718078846)
(15,2.1208533980592104)
(16,2.134868033529126)
(17,2.1468805782176026)
(18,1.7201600462572941)
(19,1.4073754900696671)
(20,1.3897665920236928)
(21,1.0234734846180238)
(22,1.0019084853685845)
(23,0.9803434861191459)
};
\addlegendentry{SCR at bus 24}
\end{axis}

% ========== 新增右侧Y轴：用电需求 Load (MWh) ==========
\begin{axis}[
    width=0.95\linewidth, height=3.5cm,
    xmin=0, xmax=23,
    ymin=160, ymax=480, % 适配数据最小值176、最大值451
    axis y line*=right,
    axis x line=none, % 隐藏底部x轴，共用左轴x刻度
    ylabel={\scriptsize \parbox[c]{2cm}{\centering Load (MWh) }}, 
    xtick=\empty, % 不重复绘制x刻度
    ytick={200,260,320,380,440},
    tick label style={font=\scriptsize},
    tick align=outside,
    ylabel style={at={(axis description cs:1.10,0.5)}},
    legend style={
        font=\scriptsize,
        fill=none,
        draw=none,
        at={(0.265,0.68)}, anchor=north east,
        row sep=-4pt,
    },
    legend cell align={left},
    legend image post style={scale=1},
    legend image code/.code={%
        \draw[mark repeat=1,mark phase=1,#1] plot coordinates {(0,0)};
    }
]
% 用电需求曲线：灰色虚线三角标记，区分SCR曲线
\addplot+[color=black!60, mark=triangle*, mark size=1.5pt, line width=0.6pt, dashed] coordinates {
(0,192.50000000000006)
(1,178.09000000000006)
(2,188.54000000000002)
(3,195.14000000000004)
(4,183.59000000000006)
(5,176.0)
(6,224.07)
(7,300.5200000000001)
(8,337.15000000000003)
(9,358.3800000000001)
(10,359.59000000000003)
(11,355.4100000000001)
(12,356.29)
(13,356.9500000000001)
(14,369.3800000000001)
(15,355.96000000000004)
(16,423.6100000000001)
(17,451.00000000000017)
(18,400.73000000000013)
(19,352.99)
(20,305.14)
(21,241.78000000000003)
(22,184.80000000000004)
(23,179.96000000000004)
};
\addlegendentry{Load profile}
\end{axis}

\end{tikzpicture}
\end{adjustbox}
\caption{SCR of buses 23 and 24, and system load over 24 hours.}
\label{fig:scr_24h}
\end{figure}

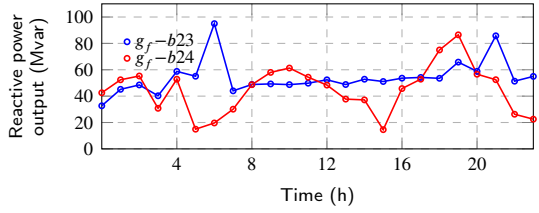
\begin{figure}[!t]
\centering
\begin{adjustbox}{minipage=[t]{0.87\linewidth},center}
\begin{tikzpicture}

% ========== 左侧Y轴：无功注入 Reactive power injection (Mvar) ==========
\begin{axis}[
    width=1\linewidth, height=3.5cm,
    xlabel={\scriptsize Time (h)},
    ylabel={\scriptsize \parbox[c]{2cm}{\centering Reactive power output (Mvar) }}, 
    label style={font=\scriptsize}, tick label style={font=\scriptsize},
    xmin=0, xmax=23,
    ymin=0, ymax=110,
    xtick={4,8,12,16,20},
    ytick={0,20,40,60,80,100},
    grid=both, grid style={dashed, gray!60},
    legend style={
        font=\scriptsize, 
        fill=none, 
        draw=none, 
        at={(0.14,0.49)}, anchor=south,
            row sep=-4pt
    },
    legend cell align={left},
    legend image post style={scale=1},
    legend image code/.code={%
        \draw[mark repeat=1,mark phase=1,#1] plot coordinates {(0,0)};
    },
    axis y line*=left,
    axis x line*=bottom,
    axis lines=box
]

% 无功 g_f-b23
\addplot+[color=blue, mark=o, mark size=1pt, line width=0.6pt] coordinates {
(0,32.66743421454163)
(1,45.166175484332754)
(2,48.49235259776505)
(3,40.282150872460086)
(4,58.75587588402604)
(5,55.08640202176107)
(6,94.97983390620794)
(7,44.009929743576414)
(8,48.83802888190196)
(9,49.26106314305509)
(10,48.73286766908949)
(11,49.81261406106134)
(12,52.37805948289685)
(13,48.81398810020418)
(14,52.83015307758838)
(15,51.098039933268105)
(16,53.619862359877125)
(17,54.17672801403093)
(18,53.51776490219123)
(19,65.79065057914715)
(20,58.80302634223929)
(21,85.75509992438266)
(22,51.337711308704)
(23,54.964527908811846)
};
\addlegendentry{$g_f\mathrm{-}b23$}

% 无功 g_f-b24
\addplot+[color=red, mark=o, mark size=1pt, line width=0.6pt] coordinates {
(0,42.54796051002249)
(1,52.39500086564426)
(2,55.283040281483004)
(3,30.768383889667323)
(4,52.85848312500168)
(5,14.928627439897124)
(6,19.64516405223605)
(7,30.09714020246285)
(8,48.700405699595755)
(9,57.99100513523376)
(10,61.30504810459662)
(11,54.35867296688122)
(12,48.36782324287579)
(13,37.73778463242919)
(14,37.139749961908834)
(15,14.57654425534189)
(16,45.73877428958712)
(17,52.86039217873024)
(18,75.02247322472455)
(19,86.49641985084433)
(20,56.624164922204606)
(21,52.4578812007347)
(22,26.27468409732205)
(23,22.528047969292103)
};
\addlegendentry{$g_f\mathrm{-}b24$}
\end{axis}
\end{tikzpicture}
\end{adjustbox}
\caption{Reactive power injection of $g_f\mathrm{-}b23$ and $g_f\mathrm{-}b24$ over 24 hours.}
\label{fig:reqctive_injection}
\end{figure}

This subsection first presents the provision of ancillary services and then analyzes the profitability of individual generating units under different pricing methods.

\subsubsection{Volumes of Voltage Stability Services}\label{Market-Cleared Voltage Stability Service Quantities}
The SCR contributions of synchronous units to bus 24 are shown in Fig.~\ref{fig:scr_node24}. Since unit $g_c\mathrm{-}b27$ is not committed, it does not contribute to improving the system SCR. In addition, the SCR contribution of $g_v\mathrm{-}b1$ is nearly complementary to those of the other generators. This is because increasing electricity demand coincides with reductions in both the active power and SCR support provided by $g_v\mathrm{-}b1$, which stem from weather variations. This would drive GFL-IBR at buses 23 and 24 to inject more active power to meet the rising demand. Additional SGs are then committed to offset the deficit in energy provision and SCR support from $g_v\mathrm{-}b1$, which maintains the voltage stability. Note that the SCR contributions of synchronous units to bus 23 are not plotted; their magnitudes are just marginally lower than those at bus 24.

By summing the SCR provision over the entire horizon, the resulting temporal evolution of bus SCR can be obtained as in Fig.~\ref{fig:scr_24h}. Overall, the SCR exhibits a trend consistent with system load, revealing a tight coupling between energy consumption and ancillary service provision. When synchronous units are required to meet energy demand, SCR enhancement emerges as a by-product. Conversely, units operating merely to uphold adequate SCR levels have to run at their minimum stable generation limits.

The reactive power outputs of the GFL-IBR are shown in Fig.~\ref{fig:reqctive_injection}. Since the SCR at bus 23 remains lower than that at bus 24 throughout the horizon, $g_f\mathrm{-}b23$ must provide greater reactive power support to enable higher active power injection without violating the voltage stability constraint. As a result, its hourly average active power injection is 83.73 MW/h, compared with 80.00 MW/h at bus 24, while its average reactive power output is 54.13 Mvar/h, versus 44.86 Mvar/h.

\subsubsection{Prices of Voltage Stability Services}\label{Market-Cleared Voltage Stability Service Prices}

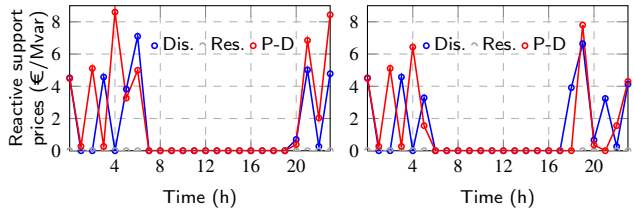
\begin{figure}[!t]
\centering
\begin{adjustbox}{minipage=[t]{1\linewidth},center}
\begin{tikzpicture}
% ====================== 左侧子图 ======================
\begin{axis}[
    width=0.6\linewidth, height=3.5cm,
    at={(-0.12\linewidth,0)}, anchor=south west, % 左子图固定画布左下角，可往左挪
    xlabel={\scriptsize Time (h)},
    ylabel={\scriptsize \parbox[c]{2cm}{\centering Reactive support prices (\texteuro/Mvar) }}, 
    ylabel shift=-5pt,
    label style={font=\scriptsize}, tick label style={font=\scriptsize},
    xmin=0, xmax=23,
    ymin=0, ymax=9,
    xtick={4,8,12,16,20},
    ytick={0,2,4,6,8},
    grid=both, grid style={dashed, gray!60},
    legend style={
        font=\scriptsize,  
        fill=none, 
        draw=none, 
        legend columns=3,
        at={(0.60,0.59)}, anchor=south,
        row sep=-4pt
    },
    legend cell align={left},
    legend image post style={scale=1},
    legend image code/.code={%
        \draw[mark repeat=1,mark phase=1,#1] plot coordinates {(0,0)};
    },
    axis y line*=left,
    axis x line*=bottom,
    axis lines=box
]
% 左子图 曲线1
\addplot+[color=blue, mark=o, mark size=1pt, line width=0.6pt] coordinates {
(0,4.5188411118597145)
(1,0)
(2,0)
(3,4.572259200326064)
(4,0)
(5,3.8242480363989317)
(6,7.100564637379762)
(7,0)
(8,0)
(9,0)
(10,0)
(11,0)
(12,0)
(13,0)
(14,0)
(15,0)
(16,0)
(17,0)
(18,0)
(19,0)
(20,0.6996804352336956)
(21,5.014991156447593)
(22,0.26733080291790984)
(23,4.7800801642457)
};
\addlegendentry{Dis.}

% 左子图 曲线2 全部置0
\addplot+[color=gray!70, mark=o, dashed, mark size=1pt, line width=0.6pt] coordinates {
(0,0) (1,0) (2,0) (3,0) (4,0) (5,0) (6,0) (7,0) (8,0) (9,0)
(10,0) (11,0) (12,0) (13,0) (14,0) (15,0) (16,0) (17,0) (18,0) (19,0)
(20,0) (21,0) (22,0) (23,0)
};
\addlegendentry{Res.}

% 左子图 曲线3
\addplot+[color=red, mark=o, mark size=1pt, line width=0.6pt] coordinates {
(0,4.485293159191423)
(1,0.2755104315482878)
(2,5.108500004190753)
(3,0.27199707194273837)
(4,8.602387604676403)
(5,3.265248263671464)
(6,4.993286944983595)
(7,0)
(8,0)
(9,0)
(10,0)
(11,0)
(12,0)
(13,0)
(14,0)
(15,0)
(16,0)
(17,0)
(18,0)
(19,0)
(20,0.3856941155511566)
(21,6.848132942319098)
(22,2.035946637781984)
(23,8.436847222433009)
};
\addlegendentry{P-D}
\end{axis}

% ====================== 右侧子图 ======================
\begin{axis}[
    width=0.6\linewidth, height=3.5cm,
    %at={(-0.3,0)}, anchor=south west, % 横向位置独立控制
    xshift=0.35\linewidth, % 控制左右子图间距
    xlabel={\scriptsize Time (h)},
    ylabel=\empty, % 取消右侧子图Y轴文字
    label style={font=\scriptsize}, tick label style={font=\scriptsize},
    xmin=0, xmax=23,
    ymin=0, ymax=9,
    xtick={4,8,12,16,20},
    ytick={0,2,4,6,8},
    grid=both, grid style={dashed, gray!60},
    legend style={
        font=\scriptsize,  
        fill=none, 
        legend columns=3,
        draw=none, 
        at={(0.50,0.59)}, anchor=south,
        row sep=-4pt
    },
    legend cell align={left},
    legend image post style={scale=1},
    legend image code/.code={%
        \draw[mark repeat=1,mark phase=1,#1] plot coordinates {(0,0)};
    },
    axis y line*=left,
    axis x line*=bottom,
    axis lines=box
]
% 右子图 曲线1
\addplot+[color=blue, mark=o, mark size=1pt, line width=0.6pt] coordinates {
(0,4.518841111907812)
(1,0)
(2,0)
(3,4.572259200303871)
(4,0)
(5,3.2862671833725337)
(6,0)
(7,0)
(8,0)
(9,0)
(10,0)
(11,0)
(12,0)
(13,0)
(14,0)
(15,0)
(16,0)
(17,0)
(18,3.916655657424446)
(19,6.641332267412239)
(20,0.6668349452227347)
(21,3.242336016816638)
(22,0.2673307881680873)
(23,4.132455264328789)
};
\addlegendentry{Dis.}

% 右子图 曲线2 全部置0
\addplot+[color=gray!70, mark=o, dashed, mark size=1pt, line width=0.6pt] coordinates {
(0,0) (1,0) (2,0) (3,0) (4,0) (5,0) (6,0) (7,0) (8,0) (9,0)
(10,0) (11,0) (12,0) (13,0) (14,0) (15,0) (16,0) (17,0) (18,0) (19,0)
(20,0) (21,0) (22,0) (23,0)
};
\addlegendentry{Res.}

% 右子图 曲线3
\addplot+[color=red, mark=o, mark size=1pt, line width=0.6pt] coordinates {
(0,4.485293159299331)
(1,0.2755104314618271)
(2,5.108500003933276)
(3,0.27199707160978104)
(4,6.434600870105088)
(5,1.5577198592292936)
(6,0)
(7,0)
(8,0)
(9,0)
(10,0)
(11,0)
(12,0)
(13,0)
(14,0)
(15,0)
(16,0)
(17,0)
(18,0)
(19,7.800151005334916)
(20,0.3621283772311076)
(21,0)
(22,1.5570973055333468)
(23,4.2988898440653305)
};
\addlegendentry{P-D}
\end{axis}
\end{tikzpicture}
\end{adjustbox}
\caption{Reactive power prices for voltage stability under different pricing methods at buses 23 (left) and 24 (right).}
\label{fig:reactive_prices_comparison}
\end{figure}

\begin{figure}[t]
\centering
\begin{adjustbox}{minipage=[t]{1\linewidth},center}
\begin{tikzpicture}
% ====================== 左侧子图 ======================
\begin{axis}[
    width=0.6\linewidth, height=3.5cm,
    at={(-0.12\linewidth,0)}, anchor=south west,
    xlabel={\scriptsize Time (h)},
    ylabel={\scriptsize \parbox[c]{2cm}{\centering Reactive support prices (\texteuro/Mvar) }}, 
    ylabel shift=-5pt,
    label style={font=\scriptsize}, tick label style={font=\scriptsize},
    xmin=0, xmax=23,
    ymin=0, ymax=9,
    xtick={4,8,12,16,20},
    ytick={0,2,4,6,8},
    grid=both, grid style={dashed, gray!60},
    legend style={
        font=\scriptsize,  
        fill=none, 
        draw=none, 
        legend columns=3,
        at={(0.60,0.59)}, anchor=south,
        row sep=-4pt
    },
    legend cell align={left},
    legend image post style={scale=1},
    legend image code/.code={%
        \draw[mark repeat=1,mark phase=1,#1] plot coordinates {(0,0)};
    },
    axis y line*=left,
    axis x line*=bottom,
    axis lines=box
]
% 左子图 Dis. 极小值<1e-5置0
\addplot+[color=blue, mark=o, mark size=1pt, line width=0.6pt] coordinates {
(0,3.4569450098492243)
(1,0)
(2,0)
(3,3.4977803643628613)
(4,0)
(5,2.6737440199217497)
(6,7.100601295622207)
(7,0)
(8,0)
(9,0)
(10,0)
(11,0)
(12,0)
(13,0)
(14,0)
(15,0)
(16,0)
(17,0)
(18,0)
(19,0)
(20,0.8651885191782775)
(21,3.614768172088477)
(22,1.314220620760557)
(23,3.6565848839915374)
};
\addlegendentry{Dis.}

% 左子图 Res. 全部极小值<1e-5置0
\addplot+[color=gray!70, mark=o, dashed, mark size=1pt, line width=0.6pt] coordinates {
(0,0)
(1,0)
(2,0)
(3,0)
(4,0)
(5,0)
(6,0)
(7,0)
(8,0)
(9,0)
(10,0)
(11,0)
(12,0)
(13,0)
(14,0)
(15,0)
(16,0)
(17,0)
(18,0)
(19,0)
(20,0)
(21,0)
(22,0)
(23,0)
};
\addlegendentry{Res.}

% 左子图 P-D 极小值<1e-5置0
\addplot+[color=red, mark=o, mark size=1pt, line width=0.6pt] coordinates {
(0,3.4311591200484104)
(1,1.3542049421859237)
(2,5.632064552846879)
(3,1.3373070076156266)
(4,6.686925929056883)
(5,3.268767678679397)
(6,4.993122523004962)
(7,0)
(8,0)
(9,0)
(10,0)
(11,0)
(12,0)
(13,0)
(14,0)
(15,0)
(16,0)
(17,0)
(18,0)
(19,0)
(20,0.8630379295612083)
(21,8.464451758908337)
(22,3.8238224638822382)
(23,6.465795472422478)
};
\addlegendentry{P-D}
\end{axis}

% ====================== 右侧子图 ======================
\begin{axis}[
    width=0.6\linewidth, height=3.5cm,
    xshift=0.35\linewidth,
    xlabel={\scriptsize Time (h)},
    ylabel=\empty,
    label style={font=\scriptsize}, tick label style={font=\scriptsize},
    xmin=0, xmax=23,
    ymin=0, ymax=9,
    xtick={4,8,12,16,20},
    ytick={0,2,4,6,8},
    grid=both, grid style={dashed, gray!60},
    legend style={
        font=\scriptsize,   
        fill=none, 
        legend columns=3,
        draw=none, 
        at={(0.50,0.59)}, anchor=south,
        row sep=-4pt
    },
    legend cell align={left},
    legend image post style={scale=1},
    legend image code/.code={%
        \draw[mark repeat=1,mark phase=1,#1] plot coordinates {(0,0)};
    },
    axis y line*=left,
    axis x line*=bottom,
    axis lines=box
]
% 右子图 Dis. 极小值<1e-5置0
\addplot+[color=blue, mark=o, mark size=1pt, line width=0.6pt] coordinates {
(0,3.456945017793582)
(1,0)
(2,0)
(3,3.4977803296237058)
(4,0)
(5,2.2977324194257935)
(6,0)
(7,0)
(8,0)
(9,0)
(10,0)
(11,0)
(12,0)
(13,0)
(14,0)
(15,0)
(16,0)
(17,0)
(18,4.277820327394131)
(19,6.434913582951152)
(20,0.8245083324707794)
(21,2.354953796317544)
(22,1.3142195297621908)
(23,3.1614718919304963)
};
\addlegendentry{Dis.}

% 右子图 Res. 全部极小值<1e-5置0
\addplot+[color=gray!70, mark=o, dashed, mark size=1pt, line width=0.6pt] coordinates {
(0,0)
(1,0)
(2,0)
(3,0)
(4,0)
(5,0)
(6,0)
(7,0)
(8,0)
(9,0)
(10,0)
(11,0)
(12,0)
(13,0)
(14,0)
(15,0)
(16,0)
(17,0)
(18,0)
(19,0)
(20,0)
(21,0)
(22,0)
(23,0)
};
\addlegendentry{Res.}

% 右子图 P-D 极小值<1e-5置0
\addplot+[color=red, mark=o, mark size=1pt, line width=0.6pt] coordinates {
(0,3.431159121117803)
(1,1.3542049408869858)
(2,5.632064548237368)
(3,1.337306996053134)
(4,5.006947527521865)
(5,1.5529998811000287)
(6,0)
(7,0)
(8,0)
(9,0)
(10,0)
(11,0)
(12,0)
(13,0)
(14,0)
(15,0)
(16,0)
(17,0)
(18,0.24280493022431382)
(19,6.959160218804705)
(20,0.8117757649155458)
(21,0)
(22,2.9101771138449584)
(23,3.280718284422895)
};
\addlegendentry{P-D}
\end{axis}
\end{tikzpicture}
\end{adjustbox}
\caption{Reactive power prices for voltage stability under different pricing methods at buses 23 (left) and 24 (right), when the startup cost is $\frac{1}{2}\cdot\mathrm{c}_{g_c}^\mathrm{st}$.}
\label{fig:reactive_prices_comparison_half_startupcost}
\end{figure}
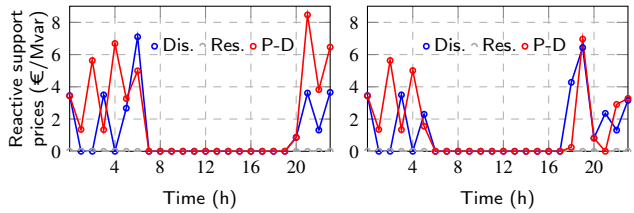

The LCOEs of VSGs and GFL-IBR are set to 12 k€/day and 13 k€/day, respectively. Note that such investment costs do not apply to the dispatchable and restricted schemes, as neither mechanism accounts for cost recovery within price formation.

The reactive support prices for voltage stability derived from various methods are illustrated in Fig.~\ref{fig:reactive_prices_comparison}. The prices of SCR enhancement are not shown because they follow the same temporal trend as reactive support prices, while remaining consistently lower. 

It can be seen that the prices of $Q_{g_f\mathrm{-}b23}$ and $Q_{g_f\mathrm{-}b24}$ under the dispatchable and P-D methods are generally zero between 06:00 and 19:00. This is because more SGs are brought online under heavy load conditions, which raises the overall system SCR. As a result, higher active power injections from GFL-IBR do not activate the voltage stability constraints. By contrast, the restricted method yields zero ancillary service prices throughout the entire horizon given this system scenario. This is due to the fact that the economic value of voltage stability support would be `absorbed' into the commitment prices. In other words, when the first-stage optimization \eqref{eq:restricted_define} determines which units are committed, the resulting commitment prices through \eqref{eq:commitment_price} already internalize the voltage stability value provided by these units and offer corresponding compensation for their fixed costs, leaving no separate remuneration for ancillary services. 

It is noteworthy that the pricing results of the dispatchable and P-D methods diverge during off-peak hours. This discrepancy stems from the P-D formulation, where the constraints \eqref{eq:nonnegative_profits_cons} bind for wind turbines, which necessitates adjustments to market clearing prices to coordinate with their service provision and guarantee no losses. This is supported by the observation that, when the LCOEs are set to zero (i.e., the wind turbines never make losses), the energy and ancillary service prices under the two methods become identical.

Furthermore, the price spikes observed in Fig.~\ref{fig:reactive_prices_comparison} are primarily associated with the fixed costs of SGs, as test results indicate that lowering startup costs can alleviate such trend, as seen in Fig.~\ref{fig:reactive_prices_comparison_half_startupcost}.

\subsubsection{Profitability of Generating Units}

\begin{table}[!t]
\caption{Market Outcome under Various Pricing Methods (k\texteuro/Day)}
\centering
\setlength{\tabcolsep}{1.2pt}
{\fontsize{8pt}{11pt}\selectfont
\begin{tabular}{ccccccccc}
\toprule
\makecell[c]{Units index}
& $g_v$-$b1$ & $g_c$-$b2$ & $g_c$-$b3$ & $g_c$-$b4$ & $g_c$-$b5$ & $g_c$-$b30$ & $g_f$-$b23$ & $g_f$-$b24$ \\
\midrule
\multicolumn{9}{c}{\textbf{Restricted pricing}} \\
\cmidrule(lr){2-9}
\makecell[c]{E. Prof.} & -2.90 & 3.06 & 2.15 & 0.08 & -0.26  & -0.96  & 5.02 & 4.60  \\
\makecell[c]{VS Rev.} & 0 & 0 & 0 & 0 & 0  & 0 & 0 & 0     \\
\makecell[c]{Commit. Rem.} & 0 & 0.57  & 0.21  & 0.30 & 0.42   & 0.96  & 0 & 0       \\
\makecell[c]{Tot. Prof.} &  \cellcolor{gray!30}{-2.90} & 3.63 & 2.36 & 0.37 & 0.15  & 0  & 5.02 & 4.60    \\
\midrule
\multicolumn{9}{c}{\textbf{Dispatchable pricing}} \\
\cmidrule(lr){2-9}
\makecell[c]{E. Prof.}   & -3.25 & 3.38 & 2.80 & 0.46 & -0.008  & -0.97 & 4.97 & 4.37 \\
\makecell[c]{VS Rev.}    & 0.66  & 0.48 & 0.11 & 0    & 0   &  1.62 & 1.97 & 1.56 \\
\makecell[c]{Tot. Prof.} & \cellcolor{gray!30}{-2.58} & 3.86 & 2.91 & 0.46 & \cellcolor{gray!30}{-0.008} &  0.65 & 6.94 & 5.93     \\
\midrule
\multicolumn{9}{c}{\textbf{Primal-dual pricing}} \\
\cmidrule(lr){2-9}
\makecell[c]{E. Prof.} & -1.90 & 4.12 & 3.35 & 0.68 & 0.15  & -0.92 & 6.92 & 6.34  \\
\makecell[c]{VS Rev.}  & 1.90  & 0.26 & 0  & 0 & 0 & 1.17  & 2.76 & 1.69      \\
\makecell[c]{Tot. Prof.} & 0  & 4.38  & 3.35   & 0.68  & 0.15  & 0.25  & 9.68 & 8.04 \\
\bottomrule
\end{tabular}
}
\label{tab:unit_revenue_dispatch_transposed}
\end{table}

Based on the market clearing results presented in Sections~\ref{Market-Cleared Voltage Stability Service Quantities} and \ref{Market-Cleared Voltage Stability Service Prices}, the settlement of individual generating units under different pricing methods is summarized in Table~\ref{tab:unit_revenue_dispatch_transposed}, in which the energy profit (E. Prof.) is calculated as energy revenue minus operating cost for thermal units, or minus LCOE for wind generators. All units except $g_v\mathrm{-}b1$ and $g_c\mathrm{-}b5$ achieve non-negative total profits (Tot. Prof.). Since $g_c\mathrm{-}b27$ is not selected to operate, its revenue remains zero and not given here.

Under restricted pricing, units $g_c\mathrm{-}b5$ and $g_c\mathrm{-}b30$ are able to offset their energy-side profit shortfalls via commitment remuneration (Commit. Rem.), thus avoiding losses. However, $g_v\mathrm{-}b1$ does not possess a commitment variable and is therefore ineligible for commitment payments. Moreover, the economic value of voltage stability is coupled only with the operating costs of thermal units and is `absorbed' by commitment prices. As a result, $g_v\mathrm{-}b1$ fails to receive explicit voltage stability service revenue (VS Rev.) to offset its LCOE-based investment cost in this system condition, leading to a negative net profit. In contrast, the negative profits of $g_v\mathrm{-}b1$ and $g_c\mathrm{-}b5$ under dispatchable pricing are simply attributable to inadequate incentives from energy and ancillary service markets.

To guarantee non-negative profits for units, the proposed P-D method jointly optimizes market clearing prices and volumes across energy and ancillary services. Consequently, $g_v\mathrm{-}b1$ and $g_c\mathrm{-}b5$ are prevented from incurring losses and have the motivation to stay in the market. Note that market clearing prices derived from the P-D method are generally higher than those from the other two pricing approaches, which delivers greater revenues to units.

\subsection{Sensitivity of Market Clearing to Electricity Demand}
This subsection investigates the impact of electricity demand levels on market clearing outcomes, including clearing prices and net profits of generating units. Note that the prices for SCR enhancement are not graphically presented.

\subsubsection{Prices vs. Electricity Demand}

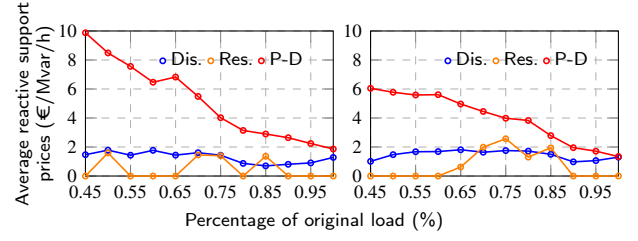
\begin{figure}[t]
\centering
\begin{adjustbox}{minipage=[t]{1\linewidth},center}
\begin{tikzpicture}
% ====================== 左侧子图 ======================
\begin{axis}[
    width=0.58\linewidth, height=3.5cm,
    at={(-0.12\linewidth,0)}, anchor=south west,
    ylabel={\scriptsize \parbox[c]{3cm}{\centering Average reactive support prices (\texteuro/Mvar/h) }}, 
    ylabel shift=-5pt,
    label style={font=\scriptsize}, tick label style={font=\scriptsize},
    xmin=0.45, xmax=1.00,
    xtick={0.45,0.55,0.65,0.75,0.85,0.95},
    xticklabel style={font=\scriptsize},
    ymin=0, ymax=10,
    ytick={0,2,4,6,8,10},
    grid=both, grid style={dashed, gray!60},
    legend style={
        font=\scriptsize,  
        fill=none, 
        draw=none, 
        legend columns=3,
        at={(0.60,0.66)}, anchor=south,
        row sep=-4pt
    },
    legend cell align={left},
    legend image post style={scale=1},
    legend image code/.code={%
        \draw[mark repeat=1,mark phase=1,#1] plot coordinates {(0,0)};
    },
    axis y line*=left,
    axis x line*=bottom,
    axis lines=box
]
% 左子图 Dis.
\addplot+[color=blue, mark=o, mark size=1pt, line width=0.6pt] coordinates {
(0.45,1.4696327063409387)
(0.50,1.7841360327547706)
(0.55,1.4340572438364105)
(0.60,1.769499857481964)
(0.65,1.4419638260113778)
(0.70,1.608698042579143)
(0.75,1.4325945228348624)
(0.80,0.8701208869443545)
(0.85,0.6990418609332698)
(0.90,0.8094485254370092)
(0.95,0.9044610237301239)
(1.00,1.2824164896720764)
};
\addlegendentry{Dis.}

% 左子图 Res. 极小值(<1e-5)全部替换为0
\addplot+[color=orange, mark=o, mark size=1pt, line width=0.6pt] coordinates {
(0.45,0)
(0.50,1.594716935881882)
(0.55,0)
(0.60,0)
(0.65,0)
(0.70,1.4511008773274634)
(0.75,1.3998288980470859)
(0.80,0)
(0.85,1.3678528104622523)
(0.90,0)
(0.95,0)
(1.00,0)
};
\addlegendentry{Res.}

% 左子图 P-D
\addplot+[color=red, mark=o, mark size=1pt, line width=0.6pt] coordinates {
(0.45,9.876662114152369)
(0.50,8.485837212778433)
(0.55,7.558327531961648)
(0.60,6.466074156039094)
(0.65,6.8237714965791625)
(0.70,5.485701668888498)
(0.75,4.022020121609635)
(0.80,3.141774671584661)
(0.85,2.898154290241434)
(0.90,2.642874547336094)
(0.95,2.2419970306915324)
(1.00,1.8630527910383317)
};
\addlegendentry{P-D}
\end{axis}

% ====================== 右侧子图 ======================
\begin{axis}[
    width=0.58\linewidth, height=3.5cm,
    xshift=0.33\linewidth,
    ylabel=\empty,
    label style={font=\tiny}, tick label style={font=\scriptsize},
    xmin=0.45, xmax=1.00,
    xtick={0.45,0.55,0.65,0.75,0.85,0.95},
    xticklabel style={font=\scriptsize},
    ymin=0, ymax=10,
    ytick={0,2,4,6,8,10},
    grid=both, grid style={dashed, gray!60},
    legend style={
        font=\scriptsize, 
        fill=none, 
        legend columns=3,
        draw=none, 
        at={(0.60,0.66)}, anchor=south,
        row sep=-4pt
    },
    legend cell align={left},
    legend image post style={scale=1},
    legend image code/.code={%
        \draw[mark repeat=1,mark phase=1,#1] plot coordinates {(0,0)};
    },
    axis y line*=left,
    axis x line*=bottom,
    axis lines=box
]
% 右子图 Dis.
\addplot+[color=blue, mark=o, mark size=1pt, line width=0.6pt] coordinates {
(0.45,1.0115385922842137)
(0.50,1.4740472714385682)
(0.55,1.674340444016887)
(0.60,1.6853571071479296)
(0.65,1.8103775645288838)
(0.70,1.6404443716462402)
(0.75,1.7547130309143586)
(0.80,1.7206077617255078)
(0.85,1.497797138945229)
(0.90,0.9767446184262534)
(0.95,1.0615654142771336)
(1.00,1.301846359990131)
};
\addlegendentry{Dis.}

% 右子图 Res. 极小值(<1e-5)全部替换为0
\addplot+[color=orange, mark=o, mark size=1pt, line width=0.6pt] coordinates {
(0.45,0)
(0.50,0)
(0.55,0)
(0.60,0)
(0.65,0.6200608959718324)
(0.70,1.9888735539153615)
(0.75,2.570279777195177)
(0.80,1.304893301450364)
(0.85,1.9405580714227835)
(0.90,0)
(0.95,0)
(1.00,0)
};
\addlegendentry{Res.}

% 右子图 P-D
\addplot+[color=red, mark=o, mark size=1pt, line width=0.6pt] coordinates {
(0.45,6.046176605205047)
(0.50,5.773913489782266)
(0.55,5.58617024225525)
(0.60,5.609631645042639)
(0.65,4.964131024495506)
(0.70,4.451612002836896)
(0.75,3.986091567567)
(0.80,3.830412669106257)
(0.85,2.7872715568321094)
(0.90,1.9538541192021672)
(0.95,1.7155321135102077)
(1.00,1.3395215287259212)
};
\addlegendentry{P-D}
\end{axis}
\node[font=\scriptsize, anchor=north, yshift=8pt] at (current bounding box.south) {Percentage of original load (\%)};
\end{tikzpicture}
\end{adjustbox}
\caption{Daily average reactive power prices for voltage stability under different pricing methods at buses 23 (left) and 24 (right), as a function of load levels.}
\label{fig:reactive_prices_comparison_loadlevels}
\end{figure}

\begin{figure}[!t]
\centering
\begin{adjustbox}{minipage=[t]{1\linewidth},center}
\begin{tikzpicture}
% ====================== 左侧子图 ======================
\begin{axis}[
    width=0.57\linewidth, height=3.5cm,
    at={(-0.12\linewidth,0)}, anchor=south west,
    ylabel={\scriptsize \parbox[c]{2.2cm}{\centering Average wind curtailment (MW/h) }}, 
    ylabel shift=-5pt,
    label style={font=\scriptsize}, tick label style={font=\scriptsize},
    xmin=0.45, xmax=1.00,
    xtick={0.45,0.55,0.65,0.75,0.85,0.95},
    xticklabel style={font=\scriptsize},
    ymin=10, ymax=40,
    ytick={10,20,30,40},
    grid=both, grid style={dashed, gray!60},
    legend style={
        font=\scriptsize, 
        fill=none, 
        draw=none, 
        legend columns=3,
        at={(0.72,0.7)}, anchor=south,
        row sep=-4pt
    },
    legend cell align={left},
    legend image post style={scale=1},
    legend image code/.code={%
        \draw[mark repeat=1,mark phase=1,#1] plot coordinates {(0,0)};
    },
    axis y line*=left,
    axis x line*=bottom,
    axis lines=box
]
% 左子图 g_f-b23
\addplot+[color=blue, mark=o, mark size=1pt, line width=0.6pt] coordinates {
(0.45,38.755338615482025)
(0.50,34.47590725678996)
(0.55,30.241489338290048)
(0.60,27.37069274628941)
(0.65,24.746255769253114)
(0.70,23.010080774261287)
(0.75,20.202079687560804)
(0.80,18.683135616929942)
(0.85,17.594604154454146)
(0.90,15.3418269989165)
(0.95,13.325134563886962)
(1.00,11.035873647226532)
};
\addlegendentry{$g_f\mathrm{-}b23$}

% 左子图 g_f-b24
\addplot+[color=red, mark=o, mark size=1pt, line width=0.6pt] coordinates {
(0.45,38.39884172795579)
(0.50,32.952007639595166)
(0.55,27.492772268599577)
(0.60,24.27322181883709)
(0.65,23.297138013521558)
(0.70,21.378210140552625)
(0.75,19.352418407171445)
(0.80,17.27616670728646)
(0.85,14.956155336228997)
(0.90,14.137696763180893)
(0.95,11.74279185408993)
(1.00,10.588095418241208)
};
\addlegendentry{$g_f\mathrm{-}b24$}
\end{axis}

% ====================== 右侧子图 ======================
\begin{axis}[
    width=0.57\linewidth, height=3.5cm,
    xshift=0.36\linewidth,
    ylabel={\scriptsize \parbox[c]{2.2cm}{\centering Average SCR }}, 
    ylabel shift=-7.5pt,
    label style={font=\scriptsize}, tick label style={font=\scriptsize},
    xmin=0.45, xmax=1.00,
    xtick={0.45,0.55,0.65,0.75,0.85,0.95},
    xticklabel style={font=\scriptsize},
    ymin=0.6, ymax=1.45,
    ytick={0.6,0.8,1.0,1.2,1.4},
    grid=both, grid style={dashed, gray!60},
    legend style={
        font=\scriptsize,  
        fill=none, 
        legend columns=3,
        draw=none, 
        at={(0.65,0.7)}, anchor=south,
        row sep=-4pt
    },
    legend cell align={left},
    legend image post style={scale=1},
    legend image code/.code={%
        \draw[mark repeat=1,mark phase=1,#1] plot coordinates {(0,0)};
    },
    axis y line*=left,
    axis x line*=bottom,
    axis lines=box
]
% 右子图 Bus 23
\addplot+[color=blue, mark=o, mark size=1pt, line width=0.6pt] coordinates {
(0.45,0.6315126992879995)
(0.50,0.6595397651017681)
(0.55,0.7335383065762825)
(0.60,0.8243478236607175)
(0.65,0.8593829771222893)
(0.70,0.864350623302801)
(0.75,0.8552487186679651)
(0.80,0.8923426237352353)
(0.85,0.9790540464022763)
(0.90,1.0693264642049412)
(0.95,1.1906785634462387)
(1.00,1.2835633415872747)
};
\addlegendentry{Bus 23}

% 右子图 Bus 24
\addplot+[color=red, mark=o, mark size=1pt, line width=0.6pt] coordinates {
(0.45,0.6775122024675081)
(0.50,0.7108639203211572)
(0.55,0.7993223254640375)
(0.60,0.9036639539035773)
(0.65,0.9305379884272719)
(0.70,0.9378416389969493)
(0.75,0.9107275992323519)
(0.80,0.9527794338837419)
(0.85,1.0515058212040882)
(0.90,1.1447834854632735)
(0.95,1.2945765474887652)
(1.00,1.4043797919579017)
};
\addlegendentry{Bus 24}
\end{axis}
% 全局共用底部X轴标签
\node[font=\scriptsize, anchor=north, yshift=1pt] at (current bounding box.south) {Percentage of original load (\%)};
\end{tikzpicture}
\end{adjustbox}
\caption{Daily average wind curtailment (left) and SCR (right) of buses 23 and 24, as a function of load levels.}
\label{fig:curtail_scr_loadlevels_compare}
\end{figure}
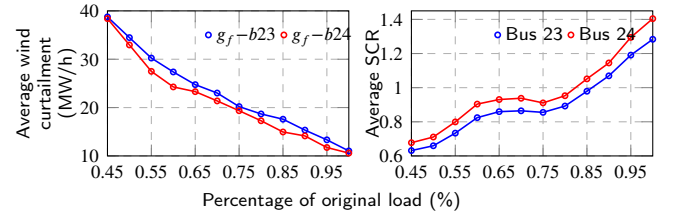

\begin{figure}[!t]
\centering
\begin{adjustbox}{minipage=[t]{1\linewidth},center}
\begin{tikzpicture}
% ========== 左侧Y轴：平均电价 ==========
\begin{axis}[
    width=0.89\linewidth, height=3.5cm,
    xmin=0.45, xmax=1.00,
    ymin=0, ymax=10,
    xlabel={\scriptsize Percentage of original load (\%)},
    xlabel style={yshift=-8pt},
    ylabel={\scriptsize \parbox[c]{2.2cm}{\centering Average energy prices (\texteuro/MW)} }, 
    xtick={0.45,0.55,0.65,0.75,0.85,0.95},
    xticklabel style={font=\scriptsize},
    ytick={0,2,4,6,8,10},
    tick label style={font=\scriptsize},
    axis y line*=left,
    axis x line*=bottom,
    axis lines=box,
    grid=both, grid style={dashed, gray!60},
    legend style={
            legend columns=3,
        font=\scriptsize, 
        fill=none,
        draw=none,
        at={(0.32,1)}, anchor=north east,
        row sep=-3pt
    },
    legend cell align={left},
    legend image post style={scale=1},
    legend image code/.code={%
        \draw[mark repeat=1,mark phase=1,#1] plot coordinates {(0,0)};
    }
]

% Dis. 电价
\addplot+[color=blue, mark=o, mark size=1pt,line width=0.6pt] coordinates {
(0.45,0.5397960433225001)
(0.50,1.1126693699569372)
(0.55,2.1940105967107146)
(0.60,3.035500708072961)
(0.65,3.44723973070692)
(0.70,3.663444590944994)
(0.75,4.1644420401893765)
(0.80,4.563433883408025)
(0.85,5.736378575197134)
(0.90,6.817449619521594)
(0.95,7.56123258591887)
(1.00,8.20828542876023)
};
\addlegendentry{Dis.}

% Res. 电价（极小值清零）
\addplot+[color=orange, mark=o, mark size=1pt, line width=0.6pt] coordinates {
(0.45,0)
(0.50,0.6399907452542181)
(0.55,2.255011086661821)
(0.60,3.958260956556676)
(0.65,2.6547383350922895)
(0.70,4.244292706759933)
(0.75,3.6629277518871035)
(0.80,3.4304222754140277)
(0.85,4.866648470951456)
(0.90,4.957461240191297)
(0.95,6.116306573469745)
(1.00,8.07569734839886)
};
\addlegendentry{Res.}

% P-D 电价
\addplot+[color=red, mark=o, mark size=1pt, line width=0.6pt] coordinates {
(0.45,3.3348116408846042)
(0.50,4.243286461074189)
(0.55,4.645183311308272)
(0.60,4.889702980987932)
(0.65,5.2367141361105904)
(0.70,5.791231566715415)
(0.75,6.563367293174836)
(0.80,7.152467224991678)
(0.85,7.931804817803798)
(0.90,8.359836814842675)
(0.95,8.787743346995276)
(1.00,9.13300644014384)
};
\addlegendentry{P-D}
\end{axis}

% ========== 右侧Y轴：机组调用总费用 ==========
\begin{axis}[
    width=0.89\linewidth, height=3.5cm,
    xmin=0.45, xmax=1.00,
    ymin=0, ymax=10,
    axis y line*=right,
    axis x line=none,
    ylabel={\scriptsize \parbox[c]{2.2cm}{\centering Total commitment payments (\texteuro)} }, 
    xtick=\empty,
    ytick={2,4,6,8,10},
    tick label style={font=\scriptsize},
    tick align=outside,
    ylabel style={at={(axis description cs:1.10,0.5)}},
    legend style={
        font=\scriptsize,  
        fill=none,
        draw=none,
        at={(0.81,1.03)}, anchor=north east,
        row sep=-4pt,
    },
    legend cell align={left},
    legend image post style={scale=1},
    legend image code/.code={%
        \draw[mark repeat=1,mark phase=1,#1] plot coordinates {(0,0)};
    }
]
% 调用费用曲线：灰色虚线三角标记
\addplot+[color=black!60, mark=triangle*, mark size=1.5pt, line width=0.6pt, dashed] coordinates {
(0.45,2.0754343295834863)
(0.50,2.704129314877735)
(0.55,3.2978188858490016)
(0.60,4.835742327883528)
(0.65,5.408013189395404)
(0.70,4.343284199413705)
(0.75,6.5166784208270565)
(0.80,6.2988293005378)
(0.85,6.596531707304227)
(0.90,7.511593799454772)
(0.95,7.534699024313251)
(1.00,6.108223671925494)
};
\addlegendentry{Commit payments ($400^{-1}$)}
\end{axis}
\end{tikzpicture}
\end{adjustbox}
\caption{Daily average energy prices under various pricing methods, as well as total commitment payments, as a function of load levels.}
\label{fig:price_commit_compare}
\end{figure}
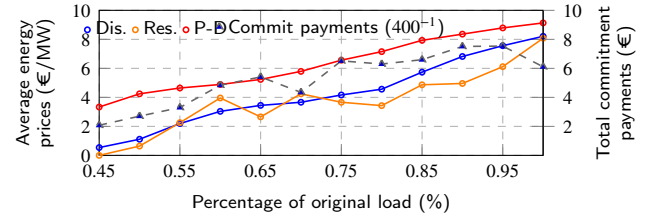

\pgfplotstableread{
X Y Z
0.45 1  0.0000
0.50 1  0.0000
0.55 1  0.0000
0.60 1  0.0000
0.65 1  0.0000
0.70 1  0.0000
0.75 1  0.0000
0.80 1  0.0000
0.85 1  0.0000
0.90 1  -137.5528
0.95 1  233.2561
1.00 1  458.9556

0.45 2  0.0000
0.50 2  0.0000
0.55 2  0.0000
0.60 2  0.0000
0.65 2  0.0000
0.70 2  0.0000
0.75 2  0.0000
0.80 2  0.0000
0.85 2  0.0000
0.90 2  92.0333
0.95 2  278.2727
1.00 2  370.6525

0.45 3  0.0000
0.50 3  0.0000
0.55 3  0.0000
0.60 3  0.0000
0.65 3  0.0000
0.70 3  0.0000
0.75 3  0.0000
0.80 3  0.0000
0.85 3  0.0000
0.90 3  62.2134
0.95 3  480.9852
1.00 3  677.7364

0.45 4  0.0000
0.50 4  0.0000
0.55 4  0.0000
0.60 4  0.0000
0.65 4  0.0000
0.70 4  0.0000
0.75 4  0.0000
0.80 4  0.0000
0.85 4  -119.0305
0.90 4  0.0000
0.95 4  -54.5351
1.00 4  -8.0981 

0.45 5  0.0000
0.50 5  0.0000
0.55 5  0.0000
0.60 5  0.0000
0.65 5  0.0000
0.70 5  0.0000
0.75 5  0.0000
0.80 5  0.0000
0.85 5  -0.1134 
0.90 5  0.0000
0.95 5  -0.3756
1.00 5  153.2063

0.45 6  0.0000
0.50 6  0.0000
0.55 6  0.0000
0.60 6  0.0000
0.65 6  0.0000
0.70 6  0.0000
0.75 6  0.0000
0.80 6  0.0000
0.85 6  0.0000
0.90 6  0.0000
0.95 6  0.0000
1.00 6  145.0169

0.45 7  -11025.9465
0.50 7  -10281.5501
0.55 7  -9514.4972
0.60 7  -8265.0035
0.65 7  -8082.6895
0.70 7  -7657.9222
0.75 7  -7068.7404
0.80 7  -6795.4310
0.85 7  -5675.7150
0.90 7  -4428.2419
0.95 7  -3539.5222
1.00 7  -2583.8683

0.45 8  -11999.8978
0.50 8  -10675.2926
0.55 8  -10051.4198
0.60 8  -8720.2173
0.65 8  -9683.6173
0.70 8  -6550.7364
0.75 8  -7039.2107
0.80 8  -8669.3970
0.85 8  -5730.3840
0.90 8  -6788.4505
0.95 8  -5614.4022
1.00 8  -2894.5952

0.45 9  0.0000
0.50 9  0.0000
0.55 9  0.0000
0.60 9  0.0000
0.65 9  0.0000
0.70 9  0.0000
0.75 9  0.0000
0.80 9  0.0000
0.85 9  0.0000
0.90 9  0.0000
0.95 9  0.0000
1.00 9  0.0000

0.45 10 -9459.4774
0.50 10 -7199.7552
0.55 10 -5720.0480
0.60 10 -2949.9263
0.65 10 -3279.3399
0.70 10 -2423.4461
0.75 10 -1177.7443
0.80 10 -1173.3777
0.85 10  1016.3544
0.90 10  3343.7065
0.95 10  5002.2445
1.00 10  6934.4835

0.45 11 -13000.0000
0.50 11 -7915.0424
0.55 11 -8172.4718
0.60 11 -4768.1819
0.65 11 -7314.5549
0.70 11 -372.8343
0.75 11 -2071.9024
0.80 11 -5491.7462
0.85 11  373.5404
0.90 11 -1974.6363
0.95 11  537.6195
1.00 11  5018.9775

0.45 12  4951.0808
0.50 12  6119.3663
0.55 12  6256.7879
0.60 12  7355.0879
0.65 12  6021.0446
0.70 12  5967.4913
0.75 12  6479.6238
0.80 12  6654.8651
0.85 12  7772.9784
0.90 12  8269.5808
0.95 12  9158.2410
1.00 12  9674.8423

0.45 13 -10498.3648
0.50 13 -8250.4593
0.55 13 -5886.4403
0.60 13 -3570.6672
0.65 13 -2798.0767
0.70 13 -2260.1222
0.75 13 -780.2592
0.80 13 -78.2735
0.85 13  2090.7942
0.90 13  3056.8238
0.95 13  4569.0678
1.00 13  5930.7500

0.45 14 -13000.0000
0.50 14 -12189.0037
0.55 14 -9081.6935
0.60 14 -4388.7331
0.65 14 -6206.6318
0.70 14  301.8662
0.75 14  1000.2013
0.80 14 -3122.6410
0.85 14  1601.3399
0.90 14 -2223.1402
0.95 14  435.1376
1.00 14  4595.8419

0.45 15  0.0000
0.50 15  1891.6384
0.55 15  2353.0060
0.60 15  5165.0308
0.65 15  3340.2836
0.70 15  4339.3466
0.75 15  6390.8478
0.80 15  7756.8754
0.85 15  7960.2299
0.90 15  6675.0271
0.95 15  7666.2184
1.00 15  8034.5388
}\profitData

\begin{figure}[t]
\centering
\begin{tikzpicture}

% 全局热力图样式宏，复刻模板写法
\pgfplotsset{
  profitHmap/.style={
    % 冷暖配色：深蓝负数，白色0，橙红正数，正负区分极强
  colormap={profitCoolWarm}{
    color(0.00) = (red!90!black)
    color(0.48) = (red!8!white)
    color(0.50) = (white)
    color(0.52) = (cyan!8!white)
    color(1.00) = (blue!90!black)
  },
    point meta min=-13000,
    point meta max=13000,
    width=0.73\columnwidth, 
    height=5cm,           
    xlabel={\scriptsize Percentage of original load (\%)},
    ylabel={\scriptsize Units \& Pricing methods},
    xlabel style={font=\scriptsize, yshift=2pt},
    ylabel style={font=\scriptsize, xshift=2pt, yshift=-2pt},
    xtick={0.45,0.55,0.65,0.75,0.85,0.95},
    xmin=0.4, xmax=1.05,
    xticklabel style={font=\scriptsize},    
    ytick={1,2,3,4,5,6,7,8,9,10,11,12,13,14,15},
    yticklabels={
        $g_c\mathrm{-}b4$~\&~Dis,
        $g_c\mathrm{-}b4$~\&~Res,
        $g_c\mathrm{-}b4$~\&~P-D,
        $g_c\mathrm{-}b5$~\&~Dis,
        $g_c\mathrm{-}b5$~\&~Res,
        $g_c\mathrm{-}b5$~\&~P-D,
        $g_v\mathrm{-}b1$~\&~Dis,
        $g_v\mathrm{-}b1$~\&~Res,
        $g_v\mathrm{-}b1$~\&~P-D,
        $g_f\mathrm{-}b23$~\&~Dis,
        $g_f\mathrm{-}b23$~\&~Res,
        $g_f\mathrm{-}b23$~\&~P-D,
        $g_f\mathrm{-}b24$~\&~Dis,
        $g_f\mathrm{-}b24$~\&~Res,
        $g_f\mathrm{-}b24$~\&~P-D
    },
    yticklabel style={font=\scriptsize},
    enlarge x limits={abs=0},
    enlarge y limits={abs=0.5},
  }
}

% 主热力图
\begin{axis}[
  profitHmap,
  name=mainPlot,
  colorbar=false,
]
\addplot[
  matrix plot*,
  point meta=explicit,
  mesh/cols=12,
  mesh/rows=15,
 ] table[x=X, y=Y, meta=Z] {\profitData};
\end{axis}

% 独立右侧颜色条（和模板布局完全一致）
\begin{axis}[
  name=cbar,
  at={(mainPlot.east)},
  anchor=west,
  xshift=0.1cm, 
  yshift=-1.0cm, 
  hide axis,
  scale only axis,
    width=0pt,
    height=4.8cm, 
  colormap={profitCoolWarm}{
    color(0.00) = (red!90!black)
    color(0.48) = (red!8!white)
    color(0.50) = (white)
    color(0.52) = (cyan!8!white)
    color(1.00) = (blue!90!black)
  },
  point meta min=-13000,
  point meta max=13000,
  colorbar,
  colorbar style={
    width=6pt,          
    height=3.3cm,       
    ytick={-13000,-8000,-3000,0,3000,8000,13000},
    yticklabel style={font=\scriptsize}, 
    ylabel={\scriptsize Net profits (\texteuro)},
    ylabel style={font=\scriptsize, yshift=3pt}, 
  },
]
\addplot[draw=none] coordinates {(0,0)};
\end{axis}
\end{tikzpicture}
\caption{Net profits of loss-making units, as a function of load levels.}
\label{fig:turbine_profit_heatmap_demand}
\end{figure}
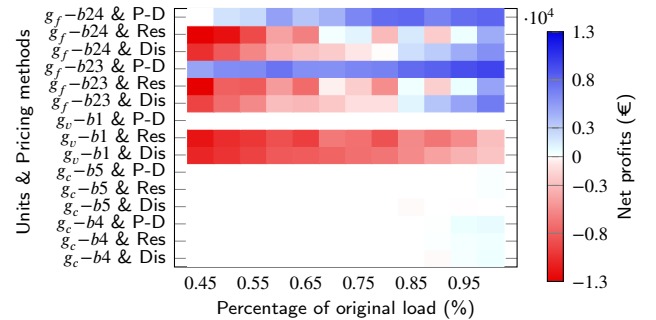

The reactive support prices under different load levels are illustrated in Fig.~\ref{fig:reactive_prices_comparison_loadlevels}. As expected, the P-D method yields significantly higher prices to guarantee non-negative profits for all units, which constitutes the only source of the pricing discrepancy compared with the dispatchable method. Moreover, because commitment prices may cover the fixed costs of SGs and compensate them for remaining online to provide voltage stability support, the explicit ancillary service prices produced by the restricted method can be zero during certain time periods.

As the load level increases, ancillary service prices under P-D formulation gradually decline. This trend can be attributed to two factors. First, wind power rises alongside growing load demand (as in the left subplot of Fig.~\ref{fig:curtail_scr_loadlevels_compare}). This pushes energy prices upward (as in Fig.~\ref{fig:price_commit_compare}), thereby enhancing energy revenues of generating units. Second, more SGs are committed for power balance, lifting the system SCR (as in the right subplot of Fig.~\ref{fig:curtail_scr_loadlevels_compare}). This relaxes the requirement for additional voltage stability support, leading generators to recover their LCOEs or operating costs primarily via the energy market and consequently suppressing ancillary service prices.

Further insights can be obtained from Fig.~\ref{fig:price_commit_compare}, which shows a complementary relationship between the energy prices and commitment payments under the restricted pricing. Specifically, beyond 60\% of the base load level, total commitment payments move inversely with energy prices, compensating for fluctuations in energy revenues. Below this load level, however, energy prices and commitment prices rise concurrently. This reveals that standalone energy prices would fail to fully compensate all thermal generators, necessitating supplementary commitment remuneration. In addition, the dispatchable method also relies on uplift payments to secure adequate generator revenues, as its resulting market prices are consistently lower than the revenue-sufficient prices derived from the P-D formulation.

Considering the tight coupling between uplift payments and energy prices, alongside full decoupling from wind generators (which carry no commitment variables), the foregoing results underscore the merits of the proposed pricing framework. By integrating cost recovery obligations into market prices, the P-D method delivers transparent and appropriate price signals independent of commitment remuneration.

\subsubsection{Net Profits vs. Electricity Demand}\label{Net Profits vs. Electricity Demand}
The profits of loss-making generators under the dispatchable and restricted schemes are visualized in Fig.~\ref{fig:turbine_profit_heatmap_demand}. It can be seen that thermal units incur very minor financial deficits (-137.55 \texteuro~at maximum) as they are hardly dispatched under light-load conditions, where wind power largely satisfies electricity demand. This very circumstance, however, causes severe losses for $g_v\mathrm{-}b1$, $g_f\mathrm{-}b23$ and $g_f\mathrm{-}b24$: market clearing prices are extremely low or even zero (as in Fig.~\ref{fig:reactive_prices_comparison_loadlevels} and Fig.~\ref{fig:price_commit_compare}), failing to cover their LCOEs. The units $g_f\mathrm{-}b23$ and $g_f\mathrm{-}b24$ can only turn profitable when high load levels allow them to earn more revenues.

By contrast, the P-D formulation delivers appropriate price signals that can be captured by wind units under any cases. These signals guarantee non-negative profits and thereby provide sufficient economic incentives for wind assets to engage in market operations.

\subsection{Sensitivity of Market Clearing to Reactive Power Support}

\begin{figure}[t]
\centering
\begin{adjustbox}{minipage=[t]{1\linewidth},center}
\begin{tikzpicture}
% ====================== 左侧子图 ======================
\begin{axis}[
    width=0.58\linewidth, height=3.5cm,
    at={(-0.12\linewidth,0)}, anchor=south west,
    ylabel={\scriptsize \parbox[c]{2.5cm}{\centering Average reactive support prices (\texteuro/Mvar/h) }}, 
    ylabel shift=-5pt,
    label style={font=\scriptsize}, tick label style={font=\scriptsize},
    xmin=0.45, xmax=1.00,
    xtick={0.45,0.55,0.65,0.75,0.85,0.95},
    xticklabel style={font=\scriptsize},
    ymin=0, ymax=3,
    ytick={0,1,2,3},
    grid=both, grid style={dashed, gray!60},
    legend style={
        font=\scriptsize, 
        fill=none, 
        draw=none, 
        legend columns=3,
        at={(0.75,0.7)}, anchor=south,
        row sep=-4pt
    },
    legend cell align={left},
    legend image post style={scale=1},
    legend image code/.code={%
        \draw[mark repeat=1,mark phase=1,#1] plot coordinates {(0,0)};
    },
    axis y line*=left,
    axis x line*=bottom,
    axis lines=box
]
% 左子图 Dis.
\addplot+[color=blue, mark=o, mark size=1pt, line width=0.6pt] coordinates {
(0.45,2.080488336140938)
(0.50,1.9568131954908399)
(0.55,1.9680755813325446)
(0.60,1.9980081959091116)
(0.65,1.4607232819032632)
(0.70,1.4759271567450905)
(0.75,1.4909883071292105)
(0.80,1.5055755586860162)
(0.85,1.4787509436825539)
(0.90,1.4870335474202088)
(0.95,1.2627862432494859)
(1.00,1.2824164896720764)
};
\addlegendentry{Dis.}

% 左子图 Res. 极小值(<1e-5)全部替换为0
\addplot+[color=orange, mark=o, mark size=1pt, line width=0.6pt] coordinates {
(0.45,0.41869834475378115)
(0.50,0)
(0.55,0)
(0.60,0)
(0.65,1.275638394879997)
(0.70,1.3118162052869755)
(0.75,1.346999291942111)
(0.80,1.9837850828354355)
(0.85,0.6161426943149488)
(0.90,0.6320144202395559)
(0.95,0.6455700731988615)
(1.00,0)
};
\addlegendentry{Res.}

% 左子图 P-D
\addplot+[color=red, mark=o, mark size=1pt, line width=0.6pt] coordinates {
(0.45,2.5150147707669555)
(0.50,2.502073745541136)
(0.55,2.5015853033618707)
(0.60,2.6254146839164534)
(0.65,2.5519760875456177)
(0.70,2.5805565293313184)
(0.75,2.425082805820954)
(0.80,2.22117626894695)
(0.85,2.109741964170516)
(0.90,1.9292204292516146)
(0.95,1.931704917143542)
(1.00,1.8630527910383317)
};
\addlegendentry{P-D}
\end{axis}

% ====================== 右侧子图 ======================
\begin{axis}[
    width=0.58\linewidth, height=3.5cm,
    xshift=0.33\linewidth,
    ylabel=\empty,
    label style={font=\scriptsize}, tick label style={font=\scriptsize},
    xmin=0.45, xmax=1.00,
    xtick={0.45,0.55,0.65,0.75,0.85,0.95},
    xticklabel style={font=\scriptsize},
    ymin=0, ymax=3,
    ytick={0,1,2,3},
    grid=both, grid style={dashed, gray!60},
    legend style={
        font=\scriptsize,  
        fill=none, 
        legend columns=3,
        draw=none, 
        at={(0.75,0.7)}, anchor=south,
        row sep=-4pt
    },
    legend cell align={left},
    legend image post style={scale=1},
    legend image code/.code={%
        \draw[mark repeat=1,mark phase=1,#1] plot coordinates {(0,0)};
    },
    axis y line*=left,
    axis x line*=bottom,
    axis lines=box
]
% 右子图 Dis.
\addplot+[color=blue, mark=o, mark size=1pt, line width=0.6pt] coordinates {
(0.45,1.6061127663279502)
(0.50,1.3836013071778162)
(0.55,1.3048908534501689)
(0.60,1.0104018510700257)
(0.65,1.2471379258969397)
(0.70,1.1676857456620895)
(0.75,1.1761397651672374)
(0.80,1.1843434870141478)
(0.85,1.1770928999791224)
(0.90,1.1850316710078894)
(0.95,1.3377450685629586)
(1.00,1.301846359990131)
};
\addlegendentry{Dis.}

% 右子图 Res. 极小值(<1e-5)全部替换为0
\addplot+[color=orange, mark=o, mark size=1pt, line width=0.6pt] coordinates {
(0.45,1.1553082780311863)
(0.50,1.57472215093786)
(0.55,0.7869898436617998)
(0.60,0)
(0.65,0.4654199498237108)
(0.70,0.4763336493041837)
(0.75,0)
(0.80,0)
(0.85,1.2100708201324997)
(0.90,0)
(0.95,0)
(1.00,0)
};
\addlegendentry{Res.}

% 右子图 P-D
\addplot+[color=red, mark=o, mark size=1pt, line width=0.6pt] coordinates {
(0.45,1.8207655523755992)
(0.50,1.7101051639709874)
(0.55,1.239181453229642)
(0.60,1.058274146581061)
(0.65,1.0318742406846082)
(0.70,0.9177680428708315)
(0.75,0.8903238499166455)
(0.80,1.1666637959165003)
(0.85,1.0761183738644988)
(0.90,1.3197154549397063)
(0.95,1.3127105904279073)
(1.00,1.3395215287259212)
};
\addlegendentry{P-D}
\end{axis}
\node[font=\scriptsize, anchor=north, yshift=3pt] at (current bounding box.south) {Percentage of installed reactive capacity (\%)};
\end{tikzpicture}
\end{adjustbox}
\caption{Daily average reactive power prices for voltage stability under different pricing methods at buses 23 (left) and 24 (right), as a function of GFL-IBR's installed reactive capacity proportion.}
\label{fig:reactive_prices_capacity_compare}
\end{figure}
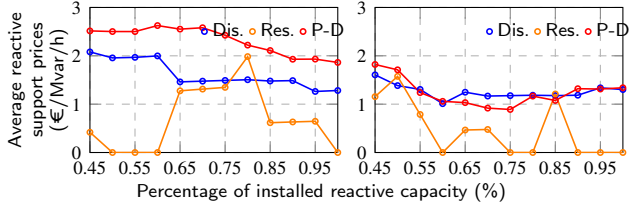

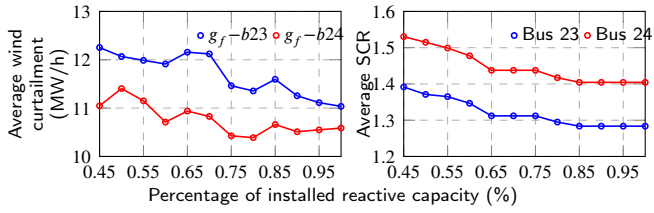
\begin{figure}[!t]
\centering
\begin{adjustbox}{minipage=[t]{1\linewidth},center}
\begin{tikzpicture}
% ====================== 左侧子图 ======================
\begin{axis}[
    width=0.57\linewidth, height=3.5cm,
    at={(-0.12\linewidth,0)}, anchor=south west,
    ylabel={\scriptsize \parbox[c]{2.2cm}{\centering Average wind curtailment (MW/h) }}, 
    ylabel shift=-5pt,
    label style={font=\scriptsize}, tick label style={font=\scriptsize},
    xmin=0.45, xmax=1.00,
    xtick={0.45,0.55,0.65,0.75,0.85,0.95},
    xticklabel style={font=\scriptsize},
    ymin=10, ymax=13,
    ytick={10,11,12,13},
    grid=both, grid style={dashed, gray!60},
    legend style={
        font=\scriptsize, 
        fill=none, 
        draw=none, 
        legend columns=3,
        at={(0.72,0.7)}, anchor=south,
        row sep=-4pt
    },
    legend cell align={left},
    legend image post style={scale=1},
    legend image code/.code={%
        \draw[mark repeat=1,mark phase=1,#1] plot coordinates {(0,0)};
    },
    axis y line*=left,
    axis x line*=bottom,
    axis lines=box
]
% 左子图 Dis.
\addplot+[color=blue, mark=o, mark size=1pt, line width=0.6pt] coordinates {
(0.45,12.252733580398248)
(0.50,12.067723167311252)
(0.55,11.987817313167218)
(0.60,11.912175914568197)
(0.65,12.156777200928474)
(0.70,12.120584471711277)
(0.75,11.462377332850316)
(0.80,11.355237047375846)
(0.85,11.595481587986669)
(0.90,11.254938394377797)
(0.95,11.113225138065213)
(1.00,11.035873647226532)
};
\addlegendentry{$g_f\mathrm{-}b23$}

% 左子图 Res.
\addplot+[color=red, mark=o, mark size=1pt, line width=0.6pt] coordinates {
(0.45,11.047362962804927)
(0.50,11.40459346109126)
(0.55,11.150418590816026)
(0.60,10.711236619967396)
(0.65,10.938997543592075)
(0.70,10.827461672465704)
(0.75,10.426423598813088)
(0.80,10.389585721093667)
(0.85,10.659793620033883)
(0.90,10.511526263346036)
(0.95,10.549995555586431)
(1.00,10.588095418241208)
};
\addlegendentry{$g_f\mathrm{-}b24$}
\end{axis}

% ====================== 右侧子图 ======================
\begin{axis}[
    width=0.57\linewidth, height=3.5cm,
    xshift=0.36\linewidth,
    ylabel={\scriptsize \parbox[c]{2.2cm}{\centering Average SCR }}, 
    ylabel shift=-7pt,
    label style={font=\scriptsize}, tick label style={font=\scriptsize},
    xmin=0.45, xmax=1.00,
    xtick={0.45,0.55,0.65,0.75,0.85,0.95},
    xticklabel style={font=\scriptsize},
    ymin=1.2, ymax=1.6,
    ytick={1.2,1.3,1.4,1.5,1.6},
    grid=both, grid style={dashed, gray!60},
    legend style={
        font=\scriptsize,  
        fill=none, 
        legend columns=3,
        draw=none, 
        at={(0.75,0.7)}, anchor=south,
        row sep=-4pt
    },
    legend cell align={left},
    legend image post style={scale=1},
    legend image code/.code={%
        \draw[mark repeat=1,mark phase=1,#1] plot coordinates {(0,0)};
    },
    axis y line*=left,
    axis x line*=bottom,
    axis lines=box
]
% 右子图 Dis.
\addplot+[color=blue, mark=o, mark size=1pt, line width=0.6pt] coordinates {
(0.45,1.3915917695705822)
(0.50,1.371150095456241)
(0.55,1.3650252566812238)
(0.60,1.3466610416816474)
(0.65,1.3119687232366826)
(0.70,1.3119687455470246)
(0.75,1.3119687232366826)
(0.80,1.294859656340283)
(0.85,1.2835633549800345)
(0.90,1.2835633415872747)
(0.95,1.2835633415872747)
(1.00,1.2835633415872747)
};
\addlegendentry{Bus 23}

% 右子图 Res.
\addplot+[color=red, mark=o, mark size=1pt, line width=0.6pt] coordinates {
(0.45,1.5303624095650272)
(0.50,1.5149003182182696)
(0.55,1.499080521971636)
(0.60,1.4776222653924036)
(0.65,1.437663181772736)
(0.70,1.4376632069243769)
(0.75,1.437663181772736)
(0.80,1.417220887787282)
(0.85,1.404379808058087)
(0.90,1.4043797919579017)
(0.95,1.4043797919579017)
(1.00,1.4043797919579017)
};
\addlegendentry{Bus 24}
\end{axis}
% 全局共用底部X轴标签
\node[font=\scriptsize, anchor=north, yshift=3pt] at (current bounding box.south) {Percentage of installed reactive capacity (\%)};
\end{tikzpicture}
\end{adjustbox}
\caption{Daily average wind curtailment (left) and SCR (right) of buses 23 and 24, as a function of GFL-IBR's installed reactive capacity proportion.}
\label{fig:curtail_scr_capacity_compare}
\end{figure}

\begin{figure}[t]
\centering
\begin{adjustbox}{minipage=[t]{1\linewidth},center}
\begin{tikzpicture}
% ========== 左侧Y轴：平均电价 ==========
\begin{axis}[
    width=0.89\linewidth, height=3.5cm,
    xmin=0.45, xmax=1.00,
    ymin=6, ymax=10,
    xlabel={\scriptsize Percentage of installed reactive capacity (\%)},
    xlabel style={yshift=-8pt},
    ylabel={\scriptsize \parbox[c]{2.2cm}{\centering Average energy prices (\texteuro/MW)} }, 
    xtick={0.45,0.55,0.65,0.75,0.85,0.95},
    xticklabel style={font=\scriptsize},
    ytick={6,7,8,9,10},
    tick label style={font=\scriptsize},
    axis y line*=left,
    axis x line*=bottom,
    axis lines=box,
    grid=both, grid style={dashed, gray!60},
    legend style={
        legend columns=3,
        font=\scriptsize, 
        fill=none,
        draw=none,
        at={(0.35,1.05)}, anchor=north east,
        row sep=-4pt
    },
    legend cell align={left},
    legend image post style={scale=1},
    legend image code/.code={%
        \draw[mark repeat=1,mark phase=1,#1] plot coordinates {(0,0)};
    }
]

% Dis. 电价
\addplot+[color=blue, mark=o, mark size=1pt,line width=0.6pt] coordinates {
(0.45,8.359029308001412)
(0.50,8.310604729556717)
(0.55,8.25004482490406)
(0.60,8.324122795530764)
(0.65,8.192038941090116)
(0.70,8.18861283637001)
(0.75,8.186947211587654)
(0.80,8.185142423694197)
(0.85,8.215685078451127)
(0.90,8.216022370261397)
(0.95,8.202523353225105)
(1.00,8.20828542876023)
};
\addlegendentry{Dis.}

% Res. 电价
\addplot+[color=orange, mark=o, mark size=1pt, line width=0.6pt] coordinates {
(0.45,7.0599199185599275)
(0.50,6.645049093748174)
(0.55,7.310235662895411)
(0.60,7.0539021639583375)
(0.65,7.694998048848173)
(0.70,7.69489128591887)
(0.75,7.694665764072084)
(0.80,7.69925570416769)
(0.85,8.073135142302421)
(0.90,8.087041549626713)
(0.95,8.07703837579981)
(1.00,8.07569734839886)
};
\addlegendentry{Res.}

% P-D 电价
\addplot+[color=red, mark=o, mark size=1pt, line width=0.6pt] coordinates {
(0.45,8.840360451746442)
(0.50,8.83110703056641)
(0.55,9.046211266054263)
(0.60,9.100661695286002)
(0.65,9.101181887545737)
(0.70,9.160965529276254)
(0.75,9.203780329374654)
(0.80,9.109534252529835)
(0.85,9.155437911434733)
(0.90,9.080517472625324)
(0.95,9.105191013442083)
(1.00,9.13300644014384)
};
\addlegendentry{P-D}
\end{axis}

% ========== 右侧Y轴：机组调用总费用 ==========
\begin{axis}[
    width=0.89\linewidth, height=3.5cm,
    xmin=0.45, xmax=1.00,
    ymin=5, ymax=8,
    axis y line*=right,
    axis x line=none,
    ylabel={\scriptsize \parbox[c]{2.2cm}{\centering Total commitment payments (\texteuro)} }, 
    xtick=\empty,
    ytick={5,6,7,8},
    tick label style={font=\scriptsize},
    tick align=outside,
    ylabel style={at={(axis description cs:1.10,0.5)}},
    legend style={
        font=\scriptsize, 
        fill=none,
        draw=none,
        at={(0.85,1.07)}, anchor=north east,
        row sep=-4pt,
    },
    legend cell align={left},
    legend image post style={scale=1},
    legend image code/.code={%
        \draw[mark repeat=1,mark phase=1,#1] plot coordinates {(0,0)};
    }
]
% 调用费用曲线：灰色虚线三角标记
\addplot+[color=black!60, mark=triangle*, mark size=1.5pt, line width=0.6pt, dashed] coordinates {
(0.45,6.692228002457476)
(0.50,6.781422935894207)
(0.55,6.909655159010303)
(0.60,7.3997185443734965)
(0.65,6.215745501974479)
(0.70,6.218939159722453)
(0.75,6.5777842418277235)
(0.80,6.072003521911566)
(0.85,5.471297534142788)
(0.90,5.731434547303151)
(0.95,5.723226516090169)
(1.00,6.108223671925494)
};
\addlegendentry{Commit payments ($400^{-1}$)}
\end{axis}
\end{tikzpicture}
\end{adjustbox}
\caption{Daily average energy prices under different pricing methods, as well as total commitment payments, as a function of GFL-IBR's installed reactive capacity proportion.}
\label{fig:price_commit_capacity_compare}
\end{figure}
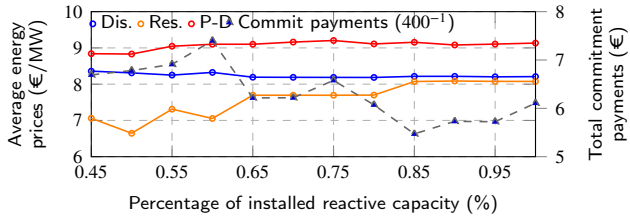

\pgfplotstableread{
X Y Z
% Y1: gc_b5-Dis
0.45 1  880.9583504690011
0.50 1  162.03821098502326
0.55 1  691.8698361220754
0.60 1  511.47129858268715
0.65 1  -28.96149043044959
0.70 1  -28.961378849541752
0.75 1  -28.961429595437355
0.80 1  -28.96075787896304
0.85 1  -17.308417365837343
0.90 1  -17.30833238324665
0.95 1  -14.63052734398626
1.00 1  -8.09814634036479

% Y2: gc_b5-Res
0.45 2  995.3511023186962
0.50 2  576.7886744673135
0.55 2  736.8819206052422
0.60 2  152.9043579850192
0.65 2  153.20610463250168
0.70 2  153.88277496169238
0.75 2  153.1515299427729
0.80 2  154.8828616800349
0.85 2  152.91732898532868
0.90 2  157.04195523821377
0.95 2  153.57255042013412
1.00 2  153.20631109745912

% Y3: gc_b5-PD
0.45 3  390.7656835849946
0.50 3  94.18722798020087
0.55 3  307.51561624714
0.60 3  303.7172794348036
0.65 3  68.43584765354515
0.70 3  94.20243027747806
0.75 3  137.35410696155205
0.80 3  124.01285518725355
0.85 3  145.01690422822116
0.90 3  145.01686993778193
0.95 3  145.01680108615045
1.00 3  145.01688313680532

% Y4: gv_b1-Dis
0.45 4  -1765.5077247163335
0.50 4  -1951.298386427121
0.55 4  -2084.0794420365182
0.60 4  -2096.805338789778
0.65 4  -2477.087770077487
0.70 4  -2488.453385080842
0.75 4  -2481.322008539668
0.80 4  -2430.9518641018076
0.85 4  -2444.175535183529
0.90 4  -2461.8568806815656
0.95 4  -2599.497007989494
1.00 4  -2583.868306368231

% Y5: gv_b1-Res
0.45 5  -4317.758325972751
0.50 5  -5017.071580019245
0.55 5  -4071.566867076247
0.60 5  -4494.591854283834
0.65 5  -2795.7058350801763
0.70 5  -2783.381262074772
0.75 5  -2771.229452435011
0.80 5  -2503.141545806875
0.85 5  -2643.4417840690694
0.90 5  -2622.5704988673115
0.95 5  -2630.0979619374266
1.00 5  -2894.595196676015

% Y6: gv_b1-PD (极小浮点全部替换为0)
0.45 6  0.0000
0.50 6  0.0000
0.55 6  0.0000
0.60 6  0.0000
0.65 6  0.0000
0.70 6  0.0000
0.75 6  0.0000
0.80 6  0.0000
0.85 6  0.0000
0.90 6  0.0000
0.95 6  0.0000
1.00 6  0.0000
}\profitData

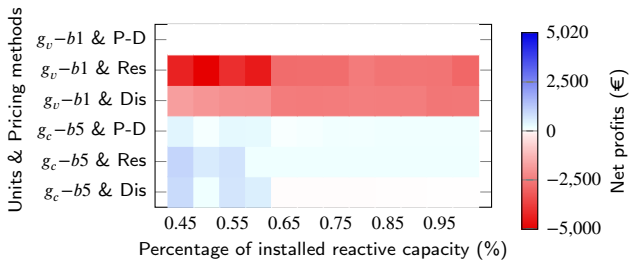
\begin{figure}[t]
\centering
\begin{tikzpicture}

% 全局热力图样式宏，修复色板非法参数，0严格纯白，负数红/正数蓝
\pgfplotsset{
  profitHmap/.style={
  colormap={profitCoolWarm}{
    color(0.00) = (red!90!black)
    color(0.48) = (red!8!white)
    color(0.50) = (white)
    color(0.52) = (cyan!8!white)
    color(1.00) = (blue!90!black)
  },
    point meta min=-5020,
    point meta max=5020,
    width=0.72\columnwidth, 
    height=4cm,           
    xlabel={\scriptsize Percentage of installed reactive capacity (\%)},
    ylabel={\scriptsize Units \& Pricing methods},
    xlabel style={font=\scriptsize, yshift=2pt},
    ylabel style={font=\scriptsize, xshift=2pt, yshift=-2pt},
    xtick={0.45,0.55,0.65,0.75,0.85,0.95},
    xmin=0.4, xmax=1.05,
    xticklabel style={font=\scriptsize},    
    ytick={1,2,3,4,5,6},
    yticklabels={
        $g_c\mathrm{-}b5$~\&~Dis,
        $g_c\mathrm{-}b5$~\&~Res,
        $g_c\mathrm{-}b5$~\&~P-D,
        $g_v\mathrm{-}b1$~\&~Dis,
        $g_v\mathrm{-}b1$~\&~Res,
        $g_v\mathrm{-}b1$~\&~P-D
    },
    yticklabel style={font=\scriptsize},
    enlarge x limits={abs=0},
    enlarge y limits={abs=0.5},
  }
}

% 主热力图
\begin{axis}[
  profitHmap,
  name=mainPlot,
  colorbar=false,
]
\addplot[
  matrix plot*,
  point meta=explicit,
  mesh/cols=12,
  mesh/rows=6,
 ] table[x=X, y=Y, meta=Z] {\profitData};
\end{axis}

% 独立右侧颜色条，同步修正色板与值域
\begin{axis}[
  name=cbar,
  at={(mainPlot.east)},
  anchor=west,
  xshift=0.1cm, 
  yshift=-1.0cm, 
  hide axis,
  scale only axis,
    width=0pt,
    height=4.2cm, 
  colormap={profitCoolWarm}{
    color(0.00) = (red!90!black)
    color(0.48) = (red!8!white)
    color(0.50) = (white)
    color(0.52) = (cyan!8!white)
    color(1.00) = (blue!90!black)
  },
  point meta min=-5020,
  point meta max=5020,
  colorbar,
  colorbar style={
    width=6pt,          
    height=2.6cm,       
    ytick={-5000,-2500,0,2500,5020},
    yticklabel style={font=\scriptsize}, 
    ylabel={\scriptsize Net profits (\texteuro)},
    ylabel style={font=\scriptsize, yshift=3pt}, 
  },
]
\addplot[draw=none] coordinates {(0,0)};
\end{axis}
\end{tikzpicture}
\caption{Net profits of loss-making units, as a function of GFL-IBR’s installed reactive capacity proportion.}
\label{fig:turbine_profit_heatmap_reactive}
\end{figure}

This subsection analyzes how reactive power support affects market clearing prices and generator net profits.

\subsubsection{Prices vs. Reactive Power Support}
The relationship between voltage-stability reactive power prices and available reactive capacity is illustrated in Fig.~\ref{fig:reactive_prices_capacity_compare}. As the available capacity increases, prices under both the P-D and dispatchable methods gradually decline. This can be explained by the voltage stability constraint, $P_{g_f}^{2} \leq \frac{1}{4}\mathrm{SCR}_{\Phi(g_f)}^{2} + Q_{g_f}\mathrm{SCR}_{\Phi(g_f)}$, under which increased reactive support can compensate for a weakened SCR and thereby facilitate greater wind power integration, as shown in Fig.~\ref{fig:curtail_scr_capacity_compare}. Consequently, both SCR enhancement and reactive support become less scarce in maintaining voltage stability, leading to lower service prices.

Overall, the prices at bus 23 remain higher than those at bus 24. This is because $g_f\mathrm{-}b23$ is connected to a weaker SCR location and therefore relies more on reactive support to accommodate large wind power injections. Accordingly, reactive power provision is more valuable at bus 23, which translates to elevated service prices. The same reasoning also applies to the pricing patterns observed in Fig.~\ref{fig:reactive_prices_comparison_loadlevels}.

Another noteworthy observation is that the ancillary service prices obtained from the P-D method at bus 24 are occasionally lower than those from the dispatchable method. This indicates that, during these periods, the market operator satisfies the profits of generating units by adjusting energy prices (as in Fig.~\ref{fig:price_commit_capacity_compare}) rather than relying on higher ancillary service prices.

As for energy prices, Fig.~\ref{fig:price_commit_capacity_compare} shows that the prices under the restricted method are generally positively correlated with the available reactive capacity. This is because smaller-capacity and higher-marginal-cost SGs more frequently serve as marginal units, driving up energy prices. Meanwhile, the corresponding commitment prices exhibit an inverse trend, indicating that the increased energy revenues are gradually able to recover a larger portion of the operating costs of thermal units, alleviating the need for commitment payments. 

This complementary relationship also explains why energy prices under the P-D and dispatchable schemes exhibit low sensitivity to reactive power capacity. The reasons are twofold. First, the fixed costs of thermal generators are already incorporated and recovered through market prices under these two mechanisms. By contrast, the restricted method relies on uplift payments to offset fluctuations in generators’ energy profits. Second, energy prices are directly tied to the capacity scarcity and marginal generation costs of thermal units, rather than the reactive power capacity of GFL-IBR.

\subsubsection{Net Profits vs. Reactive Power Support}
The impact of varying reactive power support on unit net profits is quantified as Fig.~\ref{fig:turbine_profit_heatmap_reactive}. Unit $g_c\mathrm{-}b5$ yields a profit range of \([-8.10, 28.96]\) \texteuro/day exclusively under the dispatchable scheme and generates positive earnings under the other two pricing mechanisms. Even though the resulting profit deficit is marginal (also can be seen in Section~\ref{Net Profits vs. Electricity Demand}), this observation exposes a fundamental flaw inherent to the dispatchable scheme: it cannot guarantee full cost recovery for all generators. In comparison, unit $g_v\mathrm{-}b1$ requires explicit constraints to avoid financial losses. This further demonstrates the strength of the proposed pricing framework, which is proven to provide sound economic incentives for generators independent of uplift payments.

\subsection{Computational Performance}

\begin{table}[t]
\centering
\caption{Comparison of Computational Time}
\setlength{\tabcolsep}{5pt}
{\fontsize{8pt}{8pt}\selectfont
\begin{tabular}{ll!{\vrule width 0.5pt}c}
\toprule
\multicolumn{2}{c!{\vrule width 0.5pt}}{Market clearing scenarios} & Time ($s$) \\
\midrule
\multirow{3}{*}{Energy-only market}
& Dis. (LP)   & 2.69  \\
& Res. (MILP)   & 1.94  \\
& P-D (MILP)   & 5.83  \\
\midrule
\multirow{3}{*}{\makecell[l]{Energy + voltage \\ stability service markets}}
& Dis. (SOCP)   & 4.95 \\
& Res. (MISOCP) & 3.71  \\
& P-D (MISOCP)  & 8.94 \\
\bottomrule
\end{tabular}
}
\label{table:computation_time}
\end{table}

The solving time of the market clearing in Section~\ref{Market Outcomes under Different Pricing Approaches} is listed in Table~\ref{table:computation_time}, from which it can be seen that the dispatchable and restricted methods exhibit comparable solution times and are both computationally faster than the P-D method. This is because the former two methods derive shadow prices solely through dual optimization, whereas the P-D method optimizes market prices while coordinating energy and ancillary service provision to ensure non-negative profits for all generating units, thereby increasing the computational burden inevitably. Nevertheless, compared with the clearing process of an `energy-only market', the increase in solving time remains modest, indicating that the proposed pricing method does not introduce significant computational burden.

\section{Conclusion and Future Work}\label{Conclusion}
To appropriately price voltage stability ancillary services, this paper proposes a primal-dual formulation-based pricing scheme addressing the limitation of traditional pricing methods to guarantee full cost recovery. This scheme enforces non-negative net profits for all participating units by optimizing shadow prices. Consequently, (V)SGs can earn remuneration by enhancing the SCR at weak buses, and GFL-IBR receive corresponding payments by regulating reactive power to sustain voltage stability.

Sensitivity analyses over electricity demand and reactive power support reveal that negative net profits for certain units under the dispatchable scheme stem merely from insufficient market prices, which necessitates uplift payments to cover fixed costs. While the restricted method quantifies such payments and allocates them exclusively to thermal generators, this mechanism is incompatible with IBR resources devoid of commitment variables, leaving their profitability unguaranteed and creating ambiguity regarding the allocation responsibility of uplift charges. In contrast, the primal-dual pricing framework internalizes fixed costs into transparent market prices, enabling all generating units to remain profitable without uplift payments and sustaining their market participation incentives.

Future work would incorporate stability services such as frequency reserves and inertia into the primal-dual pricing, to adequately incentivize and exploit IBR capabilities for their dominant role in future decarbonized grids. Beyond that, it is worthwhile to examine how strategic bidding by self-interested market agents affects pricing in practice.

%\section*{Acknowledgement}
%This work was supported by MICIU/AEI/10.13039/501100011033 and ERDF/EU under grant PID2023-150401OA-C22, as well as by the Madrid Government (Comunidad de Madrid-Spain) under the Multiannual Agreement 2023-2026 with Universidad Politécnica de Madrid, ``Line A - Emerging PIs''. The work of Peng Wang was also supported by China Scholarship Council under grant 202408500065.

\appendix
\section{Offline Training for Approximating Nonlinear Terms in Voltage Stability Constraints}\label{VS_cons_approx}
Let $\mathbf{z}_{g_f}=1/|Z_{\Phi({g_f})\Phi({g_f})}|$ denote the exact value obtained from the inversion of the admittance matrix, and let $z_{g_f}$ refer to its proximation, i.e., $z_{g_f} \approx \mathbf{z}_{g_f}$. Then, all feasible system operating states are enumerated, and the impedance ratios are computed from the exact network admittance matrix under the given system conditions. The candidate parameters are determined to minimize the error between the approximated and exact impedance ratios, yielding the following optimization:
\begin{subequations}\label{eq:approx_z}
\begin{align}
&     \min _{\mathcal{K}} \sum_{\omega \in \Omega}\left( z_{g_f}^{(\omega)}-\mathbf{z}_{{g_f}}^{(\omega)} \right)^{2} \label{eq:approx_min_error} \\
 & z_{g_f} = \hspace{-0.1cm}
\sum_{g_c \in \mathcal{G}_c}  \hspace{-0.1cm} k_{{g_f},g_c} u_{g_c} 
+  \hspace{-0.2cm} \sum_{g_v \in \mathcal{G}_v} \hspace{-0.1cm}  k_{{g_f},g_v} \upalpha_{g_v}  +  \hspace{-0.1cm} \sum_{m \in \mathcal{M}} \hspace{-0.1cm}  k_{{g_f},m} \eta_m \label{eq:approx_z_coefficients} 
\end{align}
\end{subequations}
where $\mathcal{K}=\{k_{{g_f},g_c}, k_{{g_f},g_v}, k_{{g_f},m}\}$ is the set of approximation coefficients, and $\upalpha_{g_v}$ denotes the historical capacity percentage of each VSG. The term $k_{g_f,m}\eta_m, m\in \mathcal{G}_c \bigcup \mathcal{G}_v$ captures interactions between each pair of (V)SGs that affect the impedance. The training dataset $\omega=\{u_{g_c}^{(\omega)},\upalpha_{g_v}^{(\omega)},\mathbf{z}_{g_f}^{(\omega)}\}\in \Omega$ is generated by enumerating all feasible (V)SG operating states.

\begin{table}[t]
\centering
\caption{Pairwise Combinations of Commitment Variables for Different Index $m$}
\setlength{\tabcolsep}{2.5pt}
{\fontsize{8pt}{12pt}\selectfont
\begin{tabular}{lccccccccccccccc}
\toprule
$m$          & 1 & 2 & 3 & 4 & 5 & 6 & 7 & 8 & 9 & 10 & 11 & 12 & 13 & 14 & 15 \\ 
\midrule
$u_{g'_c}$   & $b2$ & $b2$ & $b2$ & $b2$ & $b2$ & $b3$ & $b3$ & $b3$ & $b3$ & $b4$ & $b4$ & $b4$ & $b5$ & $b5$ & $b27$  \\ 
$u_{g''_c}$  & $b3$ & $b4$ & $b5$ & $b27$ & $b30$ & $b4$ & $b5$ & $b27$ & $b30$ & $b5$ & $b27$ & $b30$ &$b27$ &$b30$ &$b30$ \\      
\bottomrule
\end{tabular}
}
\label{table:combination_m}
\end{table}

\section{Dual Formulation of McCormick Envelopes}\label{Dual constraints for UC states in McCormick envelopes}
This section derives the term $h_{g_c,t}$ in \eqref{eq:DLL_cons_binary_stra_commit_SG_t1}-\eqref{eq:DLL_cons_binary_stra_commit_SG_t_2_T}. Taking $g_c\mathrm{-}b2$ and $g_c\mathrm{-}b3$ as an example, the composition of $h_{g_c\mathrm{-}b2,t}$ and $h_{g_c\mathrm{-}b3,t}$ is illustrated as follows:
\begin{subequations} \label{eq:h_g}
\begin{align}
& h_{g_c\mathrm{-}b2,t} = - \sum_{m=1}^{5}\gamma^{\mathrm{max}}_{1,m,t} + \sum_{m=1}^{5}\gamma^{\mathrm{min}}_{1,m,t},~ \forall t   \label{eq:eq:h_g12} \\
& h_{g_c\mathrm{-}b3,t} = - \gamma^{\mathrm{max}}_{2,1,t} \hspace{-0.1cm} + \hspace{-0.1cm} \gamma^{\mathrm{min}}_{1,1,t} \hspace{-0.1cm} - \hspace{-0.2cm}\sum_{m=6}^{9}\hspace{-0.1cm}\gamma^{\mathrm{max}}_{1,m,t} + \hspace{-0.2cm} \sum_{m=6}^{9}\hspace{-0.1cm}\gamma^{\mathrm{min}}_{1,m,t},~ \forall t \label{eq:h_g22} 
\end{align}
\end{subequations}

The dual terms for the remaining relaxed commitment variables can be formulated following the same manner. For every pairwise combination of commitment variables within set $\mathcal{M}$, four $m$-indexed auxiliary constraints, i.e., \eqref{eq:MC_linear_1}–\eqref{eq:MC_linear_3} and \eqref{eq:MC_linear_5}, need to be incorporated into the primal formulation. The total number of such combinations yields $\lvert \mathcal{M} \rvert = C_{|\mathcal{G}|}^{2} = 15$, among which the cases for $m \in [1 , 15]$ are given in Table~\ref{table:combination_m}. Further implementation details can be found in \cite{Code}.

\section{Binary Expansion for Nonlinear Terms in Non-negative Profit Constraints}\label{Linearization of Nonlinear Terms}
The bilinear product $\lambda_t^\mathrm{E} P_{g_c,t}$ is linearized via the binary expansion method, which is implemented as \cite{wang2025pricing}:
\begin{subequations}\label{eq:nonnegative_profit}
\begin{align}
& P_{g_c,t} = u_{g_c,t} (\mathrm{P}_g^\mathrm{min} + \sum_{n=1}^\mathrm{N} 2^{n-1} \Delta \mathrm{P}_{g_c} s_{n,t} ), ~ \forall t \label{eq:binary_expansion_1} \\
& \Delta \mathrm{P}_{g_c} = \frac{\mathrm{P}_{g_c}^\mathrm{max} - \mathrm{P}_{g_c}^\mathrm{min}}{2^\mathrm{N}-1},~ \forall t \label{eq:binary_expansion_2}\\
& \lambda_t^\mathrm{E} P_{g_c,t} = \lambda_t^\mathrm{E} u_{g_c,t} (\mathrm{P}_g^\mathrm{min} + \sum_{n=1}^\mathrm{N} 2^{n-1} \Delta \mathrm{P}_{g_c} s_{n,t} ),~ \forall t \label{eq:binary_expansion_3}\\
& v_{n,t} = \lambda_t^\mathrm{E} s_{n,t},~ \forall t  \label{eq:binary_expansion_4}\\
& 0 \leq \lambda_t^\mathrm{E} - v_{n,t} \leq \mathrm{M} (1-s_{n,t}),~0 \leq v_{n,t} \leq \mathrm{M} s_{n,t},~ \forall t \label{eq:binary_expansion_5}
\end{align}
\end{subequations}
where \eqref{eq:binary_expansion_1} and \eqref{eq:binary_expansion_2} decompose $P_{g_c,t}$ into a summation of auxiliary binary variables $s_{n,t}$. This reformulation rewrites the product `$\lambda_t^\mathrm{E} P_{g_c,t}$' into the form given in \eqref{eq:binary_expansion_3}. Subsequently, the bilinear term in \eqref{eq:binary_expansion_4} is linearized via constraint \eqref{eq:binary_expansion_5}, while the energy price is constrained to be non-negative to guarantee financial viability for market participants. The constant M is selected as a sufficiently large value such that \eqref{eq:binary_expansion_5} never becomes active at the optimal solution. Given that $\lambda_t^\mathrm{E}$ and $v_{n,t}$ represent energy prices, M can be set as $\mathrm{M} = 2 \cdot \mathrm{max}\{\mathrm{c}^\mathrm{m}_{g_c}~|~\forall g_c \}$. Parameter N denotes the number of discretization segments over the power output range; a larger N enables \eqref{eq:binary_expansion_1} to approximate the continuous dispatch variable $P_{g_c,t}$ with higher accuracy. 

The above linearization procedure, with appropriately selected upper bounds, can be applied analogously to the term `$(-\lambda_{2,{g_f,t}} + \mu_{g_f,t})Q_{g_f,t}$'. The product of continuous and binary variables, i.e., `$\mu_{g_f,t} u_{g_c,t}$', can be linearized in a similar manner by imposing proper bounds on $\mu_{g_f,t}$. In particular, the upper bound of $\mu_{g_f,t}$ can be set to twice the sum of the no-load and startup costs, thereby ensuring sufficient compensation for thermal units scheduled to support voltage stability. Therefore, $ \mu_{g_f,t} \leq 2 \cdot \mathrm{max} \{ \mathrm{c}_{g_c}^\mathrm{nl} + \mathrm{c}_{g_c}^\mathrm{st}~|~\forall g_c  \} $ is incorporated into the model.

\bibliographystyle{unsrt} % 按引用顺序编号

\bibliography{cas-refs}



\end{document}